\documentclass[superscriptaddress,amsmath,amssymb,pre,twocolumn]{revtex4-2}
\usepackage[margin=0.8in]{geometry}
\usepackage{algorithm}
\usepackage{algpseudocode}
\usepackage{lineno}
\usepackage{hyperref}
\usepackage{enumitem}
\usepackage{xcolor}
\usepackage[table]{xcolor}
\usepackage{amsmath, amssymb, amsthm, amsfonts} 
\usepackage{mathtools}
\usepackage{graphicx}
\usepackage{tikz}
\usetikzlibrary{arrows.meta,calc}
\graphicspath{{figures/}}
\usepackage{bbm}
\usepackage{natbib}

\DeclarePairedDelimiterX{\norm}[1]\lVert\rVert{#1}

\usepackage{wrapfig}
\usepackage[english]{babel}
\usepackage{hyperref}
\usepackage{verbatim}
\usepackage{mwe}
\usepackage{bm}
\hypersetup{
    colorlinks=true,
    citecolor=blue,
    linkcolor=blue,
    filecolor=magenta,      
    urlcolor=blue,
    pdftitle={Overleaf Example},
    pdfpagemode=FullScreen,
    }
\newcommand{\indi}{\mathbbm{1}}
\newcommand{\Ssum}{\mathop{\overset{\odot}{\sum}}\limits}
\newcommand{\err}[2]{$#1(#2)$}
\newcommand{\bv}[1]{\cellcolor{gray!18}\ensuremath{#1}}

\begin{document}
\typeout{COLUMNWIDTH=\the\columnwidth}
\typeout{TEXTWIDTH=\the\textwidth}

\title{
Percolation and clustering in ecological communities: A dynamical theory}
\author{Dario Sergo}
\affiliation{%
Statistical Physics of Computation Laboratory, \'Ecole Polytechnique F\'ed\'erale de Lausanne, Lausanne, Switzerland
}%
 \author{Cédric Koller}%
\affiliation{%
Statistical Physics of Computation Laboratory, \'Ecole Polytechnique F\'ed\'erale de Lausanne, Lausanne, Switzerland
}%
\author{Vittorio Erba}
\affiliation{%
Statistical Physics of Computation Laboratory, \'Ecole Polytechnique F\'ed\'erale de Lausanne, Lausanne, Switzerland
}%
\author{Lenka Zdeborová}%
\affiliation{%
Statistical Physics of Computation Laboratory, \'Ecole Polytechnique F\'ed\'erale de Lausanne, Lausanne, Switzerland
}%
\begin{abstract}

Ecological communities with structured interactions exhibit collective phenomena such as percolation and clustering of occupied sites. While these effects have been documented in experiments and simulations, systematic analytical understanding has remained limited.
In this paper, we develop a dynamical theory of these phenomena for competitive ecological systems defined on random interaction graphs. We introduce a discrete version of the generalized Lotka–Volterra model that preserves key macroscopic features of continuous ecological dynamics while enabling analytical treatment. Within this framework, we characterize the emergence of percolating clusters and describe the spatial organization of surviving sites.

Our analysis uncovers which equilibria can be reached by the dynamics and shows how this dynamical accessibility governs the onset of clustering and percolation. In doing so, our framework complements classical Lotka–Volterra theory by providing a dynamical perspective on the collective organization of structured communities.

\end{abstract}
\maketitle

\section*{Significance statement}
Ecological communities often exhibit large-scale patterns and undergo percolation and clustering phenomena. These collective processes have been observed both in natural ecosystems and numerical models, yet their analytical understanding has remained limited. We develop a dynamical theory that connects local competitive interactions to the large-scale organization of ecological communities. Using a solvable variant of the generalized Lotka–Volterra model, we characterize when surviving species self-organize in extensive connected clusters and when communities fragment into disconnected patches. Our results provide an analytical explanation, based on a theory that explicitly tracks dynamical transients, for how local ecological interactions generate large-scale community structure. Furthermore, beyond ecology, the theoretical framework introduced here provides a general approach for studying percolation phenomena generated by interacting dynamical systems.

\section{Introduction}

Ecological communities often display striking collective phenomena, including percolation, pattern formation, and clustering~\cite{Review,borgogno2009mathematical}. Such an organization has been extensively documented in vegetation systems and dryland ecosystems, where local facilitation and competition generate large-scale patterns and percolation-like transitions \cite{ManorStructured,dakos2011slowing}.
Furthermore, the type of spatial patterns that emerge are reportedly shaped by external environmental drivers. For example, \cite{Kefi2} reported that, in a Mediterranean ecosystem, increasing grazing pressure (the amount of vegetation consumed by herbivores in a given area) progressively reduces vegetation patch size and appears to induce a percolation-like transition.
Similarly, mathematical modeling and field observations reveal that rainfall plays an equivalent role in shaping the structure of dry ecosystems \cite{von2001diversity,rietkerk2002self}. 
Understanding how these collective structures emerge from local interactions is thus a central question in community ecology.

Lattice-based and cellular automaton simulations
have played a major role in addressing this question \cite{Paradigm,ermentrout1993cellular,BookCellular}. In these systems, communities are defined on a grid and neighboring sites interact through facilitative or competitive mechanisms \cite{Turing,d2006patterns,lejeune1999short}. At their steady states, these models can successfully reproduce experimentally observed vegetation patchiness and connectivity transitions, and have revealed a rich phase diagram \cite{Kefi1,Kefi2,vega2011effects}. 
The transient dynamics can also play an important role in shaping the spatial organization of a competitive system \cite{ge2023hidden}. 
However, the analysis of these models has largely remained numerical in nature~\cite{Kefi1,Kefi2,vega2011effects,Review, ge2023hidden}.

\begin{figure*}[!t]
    \centering
    \includegraphics[width=\textwidth]{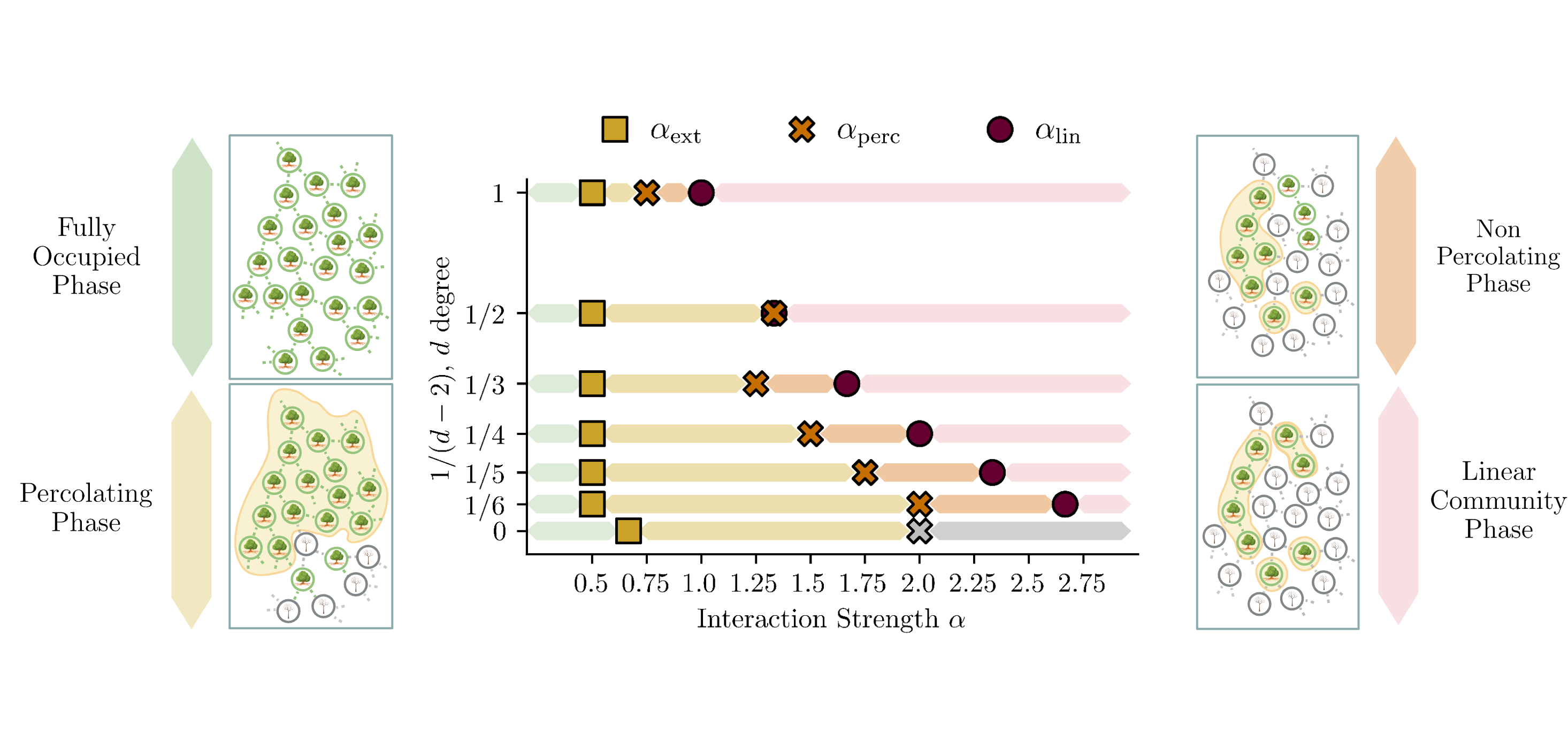}
    \caption{\textbf{Spatial phase diagram of the discrete gLV model}, on a $d$-regular graph with uniform competitive interactions. Critical values of $\alpha$ at which some biomasses become zero for some sites ($\alpha_{\rm ext}$), at which the percolation transition occurs ($\alpha_{\rm perc}$), and at which the clusters of occupied sites become linear ($\alpha_{\rm lin}$). A graphical representation of the different phases is also provided. This phase diagram is produced for the simplest instructive case with carrying capacity $K=2$ (the phenomenology remains qualitatively the same at higher $K$, see Appendix~\ref{Appendix_Numeric} and the Table~\ref{Diagram_K=3} in Appendix~\ref{Additonal_BDCM}). The critical extinction threshold in the fully connected limit ($d=\infty$) is also shown, demonstrating that the constancy of $\alpha_{\rm ext}$ does not persist as $d\gg 1$ (see Figure~\ref{high_degree_K_2} in Appendix~\ref{High_d_appendix} for more details). The gray band denotes the fact that for  $d=\infty$ at $\alpha>2$ the community is fully unoccupied, i.e. all sites are empty up to a sub-extensive fraction (see Figure~\ref{high_degree_K_2} in Appendix~\ref{Additonal_BDCM}). 
    Precise numerical values of the thresholds can be found in Table~\ref{Diagram_K=3} in Appendix~\ref{Additonal_BDCM}.
    }
    \label{fig1}
\end{figure*}

Our goal is to develop an analytical approach to characterize clustering and percolation in spatially structured competitive systems, linking large-scale collective phenomena to simple, local interaction rules.
A classic mathematical framework for describing ecological competition is the generalized Lotka-Volterra (gLV) model \cite{VOLTERRA1926,VitoVolterra1939,Grilli2016,coyte2015ecology,faust2012microbial,MacArthur}.  In this setting, spatial sites $i\in\{1,\ldots,S\}$ are each occupied by a total biomass $N_i$, where $S$ is the total number of sites. The biomasses $N_i$, which can be thought of as a number of individuals, evolve in time according to the following differential equation: 
\begin{equation}
\label{continuous_glv}
\frac{dN_i(t)}{dt} 
=
N_i(t)\left(\kappa - N_i(t) - \sum_{j\in\partial i} \alpha_{ij} N_j(t)\right),
\end{equation}
where $\partial i$ denotes the neighboring sites of site $i$ in a given topology, and $\alpha_{ij} > 0$ are the competitive couplings between sites. $\kappa$ represents the carrying capacity, i.e., the maximal biomass that a site can sustain at long times.
In this work, we will mainly focus on the case in which the $\alpha_{ij}$ are sampled from some probability distribution 
(as is common in random gLV models  \cite{Biroli1,AA,bunin2017ecological,marcus2022local,tonolo2026generalized,ValentinaRos}).
We remark that the gLV model can also be viewed as a limiting case of the MacArthur consumer–resource model (see Appendix~\ref{Appendix_CR} or \cite{MacArthur,MacArthurLotkaVolterraCR,Fant}). With  this interpretation in mind, the $\alpha_{ij}$ can be related to tunable environmental conditions; see \cite{gore2025transition} for an experimental verification of this relationship. Furthermore, recent work has shown, primarily through numerical simulations, that sparse competitive gLV systems can exhibit spatial clustering and percolation of the sites that remain occupied at the end of the dynamics, where a site $i$ is said to be occupied if $N_i>0$ \cite{marcus2022local}.
This suggests that the fragmentation phenomena reported in real ecosystems \cite{Kefi2, Review} as a function of  environmental parameters could be 
explained and understood in the setting of the gLV model.

We remark additionally that in the original Lotka-Volterra formulation \cite{VOLTERRA1926,VitoVolterra1939}  sites are interpreted as distinct species, with the interaction graph encoding non-spatial interactions between them. Related analytical works in high dimensional ecological systems adopting this (or similar) perspective include \cite{marcus2022local,Fabian,Biroli1,advani2018statistical,cui2020effect}.

In this work, we introduce a discrete variant of the gLV model that preserves the ecological interaction structure while bringing the dynamics closer in spirit to deterministic cellular automata models. We then develop a dynamical theory that allows us to study the self-organization of occupied sites in the ecological community analytically 
(at large times when initialized randomly), uncovering a rich phase diagram. 
We find that at low interaction strength, all sites of the considered topology are occupied (fully occupied phase).
At intermediate strength, vacant sites appear (i.e. sites with $N_i=0$, also called \textit{extinct}), but a finite fraction of the occupied ones form a percolating cluster (percolating phase), while at larger interaction strength, the occupied sites are fragmented into sub-extensive clusters (non-percolating phase).
Additionally, we show that competition may constrain the ``shape" of these clusters, for instance, limiting clusters to having linear topologies. 
Our results thus show that even a simple model can give rise to non-trivial spatial organization in the same spirit as what happens in real ecosystems \cite{ge2023hidden, Review}. 
The phase diagram is shown in Figure~\ref{fig1}.
Furthermore, we discuss attractors of the dynamics that are almost never reached from random initialization, but are accessible from carefully chosen initial conditions. We show that fully occupied attractors can exist even when random initializations lead to vacant sites, which is a result of direct relevance for restoration ecology \cite{silliman2024harnessing,orth2020restoration,wells2024seeding,corbin2012applied,grafnings2023spatial,kjaer2024spatial}.

Our theory hinges on recent advances in the analysis of dynamical systems on sparse tree-like topologies.
We first deploy the backtracking dynamical cavity method (BDCM) introduced in \cite{BDCM,CellAuto} to study properties of dominant dynamical attractors of the discrete gLV model, i.e., steady-states that are reached with high probability when the system is randomly initialized.
We then extend the BDCM analysis with a message-passing scheme to study percolation and clustering throughout the dynamics, and in particular at the steady state.
To our knowledge, this is the first time that percolation transitions have been generically characterized in highly correlated systems with non-independent edge-occupation probability; 
see \cite{PostioningNetwrok2, timonin2019statistics} instead for model-specific approaches, or \cite{Site_percolation,karrer2010message,PercolationBP,newman2002spread} for the case of independent edges/sites.
This method opens the door to studying percolation in a wide variety of systems, ranging from opinion dynamics and social systems \cite{cohen2000resilience, CellAuto,xie2022indirect} to epidemic spreading \cite{newman2002spread,karrer2010message,shrestha2015message}.

\section{The model and its phase diagram}\label{sec:model}
\textbf{Discrete gLV model.} 
We consider a set of spatial locations $i\in\{1,\dots, S\}$, distributed over a graph $G$, each occupied by 
a discrete biomass $N_i \in \mathbb{N}$.
We denote by $\underline{N}_i = (N_i^{t=1}, N_i^{t=2}, \dots)$ the dynamical trajectory of site $i$, by $\mathbf{N} = \{N_i\}_{i=1}^S$ a general system's state (i.e. biomasses configuration), and by $\mathbf{N}^t = \{ N_i^{t} \}_{i=1}^S$ the state at time $t$.
Conversely from the continuous gLV model in Eq.~\eqref{continuous_glv}, the biomasses now evolve in discrete time steps in accordance to the dynamical map $\mathbf{N}^{t+1} = \mathcal{F}(\mathbf{N}^t)$ with
\begin{equation}
\label{discrete_glv}
 \mathcal{F}(\mathbf{N})_i =
 \begin{cases}
 N_i + 1 & \text{if } g(N_i,\{N_j\}_{j\in\partial i}) > 0 \\
 N_i     & \text{if } g(N_i,\{N_j\}_{j\in\partial i}) = 0 \\
 N_i - 1 & \text{if } g(N_i,\{N_j\}_{j\in\partial i}) < 0
 \end{cases},
\end{equation}
where we introduced the \textit{growth rate} 
\begin{equation}\label{growthrate}
    g(N_i,\{N_j\}_{j\in\partial i})=N_i(K-N_i-\Sigma_{j\in\partial i} \, \, \alpha_{ij}N_j) \, ,
\end{equation}
and where $\{N_j\}_{j\in\partial i}$ denotes the set of biomasses of the neighbors of site $i$.
Here $\alpha_{ij} > 0$ are again competitive coupling strengths, and $K \in \mathbb{N}$ plays the role of a carrying capacity, similarly to $\kappa$ in Eq.~\eqref{continuous_glv}.
We assume that the initial condition is taken uniformly at random among those such that $1 \leq N_i^{t=1} \leq K$ for all sites $i$. This guarantees that $0 \leq N_i^{t} \leq K$ for all $i$ and all subsequent times $t>1$ (see Appendix~\ref{App_proof}), meaning that $K$ acts as an effective discretization parameter (the $N_i$ can take $K+1$ distinct values). Notice that we restrict our attention to trajectories for which, at the beginning of the dynamics, all sites are occupied by a non-zero biomass. This choice is made because the state $N_i=0$ is absorbing: if $N_i^{t^*}=0$, then $N_i^{t}=0$ for all $t \geq t^*$. Consequently, allowing sites to be initially \textit{vacant} (i.e. $N_i=0$) would alter the intended topology of the community, effectively restricting it to the subgraph induced by the initially occupied sites. We note that previous works also considered discretized biomasses $N_i$ while the update of the state of different sites was asynchronous and stochastic, controlled by a certain underlying stochastic process, see e.g.,  \cite{dobrinevski2012extinction,knebel2015evolutionary,Fisher}. In our model, Eq.~\eqref{discrete_glv}, the synchronous update is crucial for the analytical tractability (see later), and this is the main motivation behind it.  

In this work, we focus on interaction topologies given by a uniformly sampled random $d$-regular graph, i.e., each site is connected to exactly $d$ neighbors, and with uniform interactions, i.e., $\alpha_{ij}=\alpha/d$ (for some $\alpha>0$) for all $j\in\partial i$. We will see that this minimal ``homogeneous" setting (uniform topology, uniform interaction) is already sufficient to generate nontrivial spatial organization, such as clustering and percolation.
We show numerically in Appendix \ref{Appendix_Numeric} that these results are qualitatively robust when relaxing the topology 
(e.g., taking finite-dimensional lattices),
and when taking non-uniform competitive couplings. For instance, when simulated on a grid, our model reproduces several of the spatial patterns reported for lattices in the literature \cite{ge2023hidden,vega2011effects,Kefi1,Kefi2,BookCellular,Paradigm,ermentrout1993cellular,d2006patterns}, e.g., we recover that occupied sites form similar circular patterns as the ones described in \cite{ge2023hidden}, as well as the percolation and fragmentation phenomena observed in \cite{Kefi2,vega2011effects}. Additionally, we show that, even for moderate values of $K$ (i.e., $K=2,3$), the discrete system behaves qualitatively similarly to the continuous model in Eq.~\eqref{continuous_glv}, showing that it retains the essential features of the standard gLV dynamics (see Figure~\ref{comparison_d_3} in Appendix~\ref{Appendix_Numeric}).

We also remark that in the case of $d$-regular graphs with uniform interactions, different values of $\alpha$ can lead to the same dynamics. This is due to the discrete nature of the growth rate in Eq.~\eqref{growthrate}, and it allows us to obtain a complete characterization of the system by probing only a small subset of values of $\alpha$ (see Appendix~\ref{App_proof} for more details).

\textbf{Attractors and basins.}
Our analysis focuses on the long-time behavior of the system
in the thermodynamic limit $S \to +\infty$, and more precisely on the properties of its dynamical attractors.
A dynamical attractor of length $c$ (or $c$-cycle) of the dynamics in Eq.~\eqref{discrete_glv}  is a periodic sequence of $c$ system states, i.e. $\mathcal{F}(\mathbf{N}^{t}) = \textbf{N}^{t+1} \text{ for } t=1, \dots, c-1 \text{ and } \mathcal{F}(\mathbf{N}^{c}) = N^{1}$ (for our setting we show in Appendix~\ref{App_proof} that only attractors with $c=1,2$ exist).
We call the basin of attraction of a dynamical attractor the set of initial conditions that converge 
after an arbitrary number of time steps
to that attractor.

We will mainly focus on \textit{dominant} attractors, i.e., attractors that are sampled by initializing the system uniformly at random with $1\leq N_i^{t=1} \leq K$ (we will call this a \textit{typical initialization}), and running the dynamics to convergence.
We also define the \textit{most numerous} attractors, i.e. 
attractors that are sampled from the uniform probability distribution over all possible attractors.
In both cases, for $S\to\infty$, the properties of such attractors concentrate around their typical value (as usually happens in the thermodynamic limit). Then dominant attractors are those with the largest basins of attraction, while the most numerous attractors are those that are largest in raw number.
We stress that the typical properties of dominant and most numerous are, in general, different. Indeed, the most numerous attractors may have very small basins of attraction.

\textbf{Characterizing the structure of the system.} 
We will characterize the structure of a configuration of the system using the following observables.
 
\textit{1) Fraction of vacant sites.}
This is the fraction of sites $i^*$ such that $N_{i^*}=0$, i.e.
\begin{equation}\label{def_rho_0}
    \rho_0(\mathbf{N})=\frac{1}{S}\sum_{i=1}^S\mathbbm{1}\bigl[N_i=0\bigr],
\end{equation}
where $\mathbbm{1}$ is the indicator function.
The density $\rho_0$ is also referred to as the extinction fraction \cite{bunin2017ecological,Biroli1,marcus2022local}, when one interprets each site as a different species. 

\textit{2) Fraction of sites in the largest connected component.}
Given a topology and a configuration $\mathbf{N}$, we call \textit{cluster} any maximal connected set of occupied sites. In the biological literature, such an object is often referred to as a \textit{vegetation patch} \cite{Kefi1,Kefi2,dakos2011slowing}. The largest connected component (LC) is then the largest cluster.
We denote by $S_{LC}(\mathbf{N})$ the number of sites belonging to it, and define the corresponding fraction
\begin{equation}
    \phi_{LC}(\mathbf{N})=S_{LC}(\mathbf{N}) / S.
\end{equation}
If $\phi_{LC}\to 0$ as $S\to\infty$, then all clusters are sub-extensive and the system is in a non-percolating phase \cite{PercoBook,erd6s1960evolution}. Conversely, if $\phi_{LC}>0$ as $S\to\infty$, then a finite fraction of sites belongs to the largest cluster, and the system is percolating.

\textit{3) Structure functions.}
We define the structure function $\eta_l(\mathbf{N})$ as the fraction of occupied sites that have exactly $0\leq l\leq d$ occupied neighbors:
\begin{equation}\label{def_eta_l}
\eta_l(\mathbf{N})=
\frac{\sum_{i=1}^S
\mathbbm{1}\!\left[
\sum_{j\in\partial i}
\mathbbm{1}[N_j>0]=l
\right] \mathbbm{1}[N_i>0]}{S(1-\rho_0(\mathbf N))}
.
\end{equation}
For example, if all occupied sites are organized along \textit{linear clusters} (i.e., each site in the cluster has at most two occupied neighbors), then $\eta_l(\mathbf{N}) = 0$ for all $l \geq 3$, and if all occupied sites are isolated, then $\eta_l(\mathbf{N}) = 0$ for all $l \geq 1$. 

The observables above can be evaluated at any point of the dynamical trajectory, but in this work we are only going to focus on their value at convergence, i.e. when an attractor is reached. 
Notice that all the observables we defined are constant across the states of any dynamical attractor, since the state $N_i=0$ is absorbing and the observables depend only on whether sites are occupied or not. Thus, we extend the definition of $\rho_0, \phi_{LC}, \eta_l$ to dynamical attractors simply by computing them in any state of said attractor.

\textbf{Summary of the phase diagram of the discrete gLV model.}
We now describe the dominant dynamical attractors as a function of the interaction parameter $\alpha$ and the degree $d$ of the $d$-regular interaction topology for $S \gg 1$. 
The general picture that emerges is shown in Figure~\ref{fig1} for the carrying capacity $K=2$: as $\alpha$ is varied at fixed $d$, the system undergoes a sequence of phase transitions. 
This remains qualitatively the same for different values of $K$ (see Figure~\ref{comparison_d_3} in Appendix~\ref{Appendix_Numeric} and Table~\ref{Diagram_K=3} in Appendix~\ref{Additonal_BDCM}).
At convergence, the system will be in one of the following phases.

\textit{1) Fully occupied phase.} For $0 < \alpha < \alpha_{\rm ext}(d)$, all sites (up to a sub-extensive fraction) are occupied. To identify this phase, we compute the fraction of vacant sites $\rho_0$ for the dominant attractor, and determine the first value of $\alpha$ for which $\rho_0>0$. In a real ecosystem, this is the case in which each spatial location is occupied by some biomass, for instance, a forest with uniform vegetation coverage \cite{klausmeier1999regular23}. 
We observe that at analytically-accessible values of $3 \leq d \leq 8$ the transition value $\alpha_{\rm ext}(d)$ is constant in $d$. This is an effect of the discretization, and this behavior does not persist at larger degrees, as we show in Figure~\ref{high_degree_K_2} in Appendix~\ref{High_d_appendix} 
and by computing analytically the threshold in the $d \gg 1$ limit.

\textit{2) Percolating phase.} For $ \alpha_{\rm ext}(d) < \alpha < \alpha_{\rm perc}(d)$, the system contains a non-zero fraction of vacant sites, but is in a percolating phase. This means that a finite fraction of sites is occupied and belongs to the largest cluster. To identify this phase, we compute the fraction of sites in the largest connected component $\phi_{LC}$, and check up to which value of $\alpha$ it is non-zero. We observe that the threshold $\alpha_{\rm perc}(d)$ is monotonically increasing with $d$. This is expected, as for large $d$ all sites become more and more adjacent, implying that the largest connected component spans most of the occupied sites. Thus, at $d=\infty$ we say that the system is in a percolating phase provided that $\rho_0<1$.

\textit{3) Non-percolating phase.} For $\alpha > \alpha_{\rm perc}(d) $, the largest connected cluster of occupied sites becomes sub-extensive in size. When the system is in the non-percolating phase, the occupied sites are organized into small spatially separated clusters, corresponding qualitatively to vegetation ``patches'' \cite{ManorStructured,dakos2011slowing,Kefi2}. In the $d=\infty$ case, there is no direct equivalent of this phase. Still, in the phase diagram we denote the value of $\alpha$ at which the community becomes fully extinct, i.e. the value above which $\rho_0=1$. Above this threshold, only a sub-extensive fraction of sites has positive biomass in the community (see Appendix~\ref{Additonal_BDCM} for more details).

\textit{4) Linear community phase.} In the non-percolating phase, we identify an additional threshold $\alpha_{\rm lin}(d)$ such that, for $\alpha > \alpha_{\rm  lin}(d)$, all the clusters become linear. 
We identify this threshold by computing the value of $\alpha$ above which the structure functions $\eta_l = 0$ for all $l\geq 3$. Indeed, in a $d$-regular graph there are only $O(1)$ short loops, and the existence of circular structures can be neglected when $S\gg 1$. This is an example of a genuine spatial pattern reproduced by the model, and while real (lattice) ecosystems usually display more complex patterns, such as rings \cite{ge2023hidden} or stripes \cite{Review}, it is still interesting that even a very simple graph model presents the same phenomenology. We additionally remark, as shown in Figure~\ref{fig1} for the case $K=2$, that for some degrees $\alpha_{\rm lin}(d)$ and $\alpha_{\rm perc}(d)$ coincide, meaning that when the community fragments, the clusters are immediately linear.

\vspace{-3mm}
\section{Analytical methodology}

\subsection{Dynamical theory of extinction and topological organization} 
We now sketch the technical framework, based on \cite{BDCM, CellAuto}, that we use to study the behavior of the dynamical system of Eq.~\eqref{discrete_glv} analytically at convergence. A detailed explanation is given in Appendix~\ref{BP_theory_appendix}. We remark that this framework could be applied to general locally tree-like topologies (i.e., the topology graph $G$ has no loop of length $O(1)$ for $S \gg 1$) and to non-homogeneous coupling strengths $\alpha_{ij}$.
Here we will focus on $d$--regular topologies and uniform interactions.

Following \cite{BDCM}, we define $(p/c)$-backtracking attractors as dynamical trajectories of total length $p+c$ that, after a transient of $p$ steps, enter a cycle of length $c$. We note that at each time step the biomass $N_i$ gets updated depending only on the neighboring biomasses $\{N_j\}_{j\in\partial i}$,
 and $N_i$ itself.  In other words, the dynamical update rule is local. 
To study $(p/c)$-backtracking attractors, we introduce the uniform probability measure over fully-occupied initial conditions that, after $p\geq1$ steps of the dynamics, enter a cycle of length~$c$
\begin{equation}
\label{longequation}
\begin{aligned}
&\mathbbm{P}(\{\underline{N}_i\}_{i=1}^S)
= \frac{1}{\mathcal{Z}_{(p/c)}(G,\alpha_{ij})}\prod_{i=1}^{S} \bigg[ \mathbbm{1}\!\left[{N_i^{t=1}>0}
\right]\times
\\
&\times 
\mathbbm{1}\!\left[
N_i^{p+1}\!\!=\mathcal{F}_i\!\left(\mathbf{N}^{p+c}\right)
\right]
 \prod_{t=1}^{p+c-1}\! \!
\mathbbm{1}\!\left[
N_i^{t+1}=\mathcal{F}_i\!\left(\mathbf{N}^t\right)
\right] \bigg]
\\&
\equiv
\frac{1}{\mathcal{Z}_{(p/c)}} \prod_{i=1}^S \mathcal{A}(\underline{N}_i,\{\underline{N}_j\}_{j \in \partial i})
\, .
\end{aligned}
\end{equation}
In the last line, we introduced the shorthand $\mathcal{A}(\underline{N}_i,\{\underline{N}_j\}_{j \in \partial i})$ for the dynamical constraints, depending on a site trajectory $\underline{N}_i$ and that of its $d$ neighbors $\{\underline{N}_j\}_{j \in \partial i}$, given the locality of the update rule. 
The partition function $\mathcal{Z}_{(p/c)}(G,\alpha_{ij})$ counts the number of initial conditions that lead to a $(p/c)$-backtracking attractor, thus providing a lower bound on the size of the basins of attraction of length $c$.
Note that here we restrict the initial conditions to strictly positive biomass values, as otherwise the underlying topology of interaction would not be the graph $G$, but rather the graph restricted to occupied sites. From now on, we are going to drop the explicit dependency of $\mathcal{Z}_{(p/c)}$ on $G$ and $\alpha_{ij}$. 

We then introduce the entropy density for the size of the basin of attraction for the $(p/c)$-attractor
\begin{equation}
\Phi_{(p/c)}=\frac{1}{S}\log\bigl(\mathcal{Z}_{(p/c)}\bigr),
\end{equation}
which allows to access the average values of the observables $\rho_0$ and $\eta_{l}$ with respect to the measure in Eq.~\eqref{longequation}. 
To this end, we adopt the standard statistical physics procedure of tilting the probability measure by an exponential weight $e^{\lambda\Xi(\mathbf{N})}$, where $\lambda$ is a temperature-like parameter. Here, $\Xi(\mathbf{N})$ is a local observable that can be decomposed as a sum of site contributions, $\Xi(\mathbf{N}) = \sum_{i=1}^{S}\Xi_i(\underline{N}_i,\{\underline{N}_j\}_{j\in\partial i})$, as is the case for $\rho_0$ (Eq.~\eqref{def_rho_0}) and $\eta_l$ (Eq.~\eqref{def_eta_l}). Then the average value of $\Xi(\mathbf{N})$ over Eq.~\eqref{longequation} is given by
\begin{equation}\label{eq7}
    \langle \Xi\rangle=\frac{\partial \Phi_{(p/c)}}{\partial \lambda}\Biggl|_{\lambda=0} \, .
\end{equation}
In practice, computing the entropy for a general graph analytically is not feasible, as it is a sum over $O(K^S)$ many terms.
In the case of a locally tree-like graph $G$ and a local update rule $\mathcal{F}$, we can bypass this difficulty by using Belief Propagation (in the BDCM form \cite{CellAuto,BDCM, InfoPhysComp}) to compute the Bethe approximation of $\Phi_{(p/c)}$ and the averages of local observables. 
This method reduces the computation to $O(dK^{p+c})$ operations, in the specific case of a $d$-regular graph $G$ with uniform interaction strengths, and under the replica symmetric Ansatz (see Appendix~\ref{BP_theory_appendix}).

At the end of the derivation, given in detail in Appendix~\ref{BP_theory_appendix}, we obtain the following expressions for the entropy and the average values of the local observables (at $\lambda=0$). We stress again that all the quantities obtained with this methodology are correct provided that the replica symmetric assumption is valid:
\begin{align}
    \Phi&=\log(Z_{\rm fac})-\frac{d}{2}\log(Z_{\rm var}), \label{main:begin}\\
    Z_{\rm fac}&=\sum_{\underline{N},\{\underline{M}_i\}_{i=1}^d}\mathcal{A}(\underline{N},\{\underline{M}_i\}_{i=1}^d)\prod_{i=1}^d\chi_{\underline{M}_i,\underline{N}}^{\rightarrow},\\
 Z_{\rm var} & = \sum_{\underline{N},\underline{M}}\chi_{\underline{N},\underline{M}}^{\to}\chi_{\underline{M},\underline{N}}^{\to},\\
 \!\!\!\!\!  \langle \Xi\rangle & =
 \hspace{-5mm}
\sum_{\underline{N},\{\underline{M}_i\}_{i=1}^d}  \hspace{-4mm}
\frac{ \Xi(\underline{N},\{\underline{M}_i\}_{i=1}^d)}{Z_{\rm fac}} 
\mathcal{A}(\underline{N},\{\underline{M}_i\}_{i=1}^d)
\prod_{i=1}^d
\chi^{\to}_{\underline{M}_i,\underline{N}} \nonumber
 \end{align}
where all sums over $\underline{M}$ or $\underline{N}$ are over $\{0, \dots, K\}^{p+c}$, and $\mathcal{A}(\underline{N},\{\underline{M}_i\}_{i=1}^d)$ is defined in Eq.~\eqref{longequation}. The messages $\chi$ are obtained by solving the self-consistent equation
\begin{equation}\label{uniform_messages}
   \chi_{\underline{N},\underline{M}}^{\rightarrow}=\frac{1}{Z^{\rightarrow}}\sum_{\{\underline{M}_i\}_{i=1}^{d-1}}{\mathcal{A}(\underline{N}, \{\underline{M}_i\}_{i=1}^{d-1} \cup \underline{M})\prod_{i=1}^{d-1}\chi_{\underline{M}_i,\underline{N}}^{\rightarrow}},
\end{equation}
where $Z^{\rightarrow}$ is the normalization such that $\sum_{\underline{N},\underline{M}} \chi_{\underline{N},\underline{M}}^{\rightarrow} = 1$. 
\subsection{Theory of dynamics-dependent percolation} 
The size of the largest cluster $\phi_{LC}$ is not directly accessible from Eq.~\eqref{eq7}, since $\phi_{LC}$ is a non-local observable contrary to $\rho_0$ and $\eta_\ell$, and thus it cannot be written as a sum over sites. Moreover, the probability that multiple neighbors of a site  are  occupied (which we will call edge-occupation probability) is not factorized, featuring dynamics-dependent correlations. This is in clear contrast with the more classical edge/site-percolation framework \cite{PercolationBP,cohen2000resilience,newman2002spread,karrer2010message,Site_percolation}, where each edge/site in the underlying graph is present independently from the others.
We therefore develop an asymptotically exact message-passing method to study percolation of a network induced by an attractor of the gLV dynamics, where edge-occupation probabilities are correlated through the dynamics. Related types of dependent percolation have been considered in \cite{xie2022indirect,PostioningNetwrok2,timonin2019statistics}.
We note that our method may be applied to percolation on any dynamics-dependent system that has a probability measure that can be written in the form of Eq.~\eqref{longequation}, including epidemic models~\cite{karrer2010message,newman2002spread,newman2005threshold}, cellular automata~\cite{CellAuto,BDCM}, and  dynamics relevant in social sciences~\cite{cohen2000resilience,PostioningNetwrok2,xie2022indirect}.

In order to compute the fraction of sites in the largest component  we generalize the approach in~\cite{PercolationBP}, developed to study bond percolation. We consider the probability distribution that a site $i$ belongs to a sub-extensive cluster of $s = O(1)$ sites, conditioned on the full state of site $i$ and its neighbors.
In the gLV system, this amounts to the set of dynamical trajectories $\{\underline{N}_i, \{\underline{N}_j\}_{j \in \partial i} \}$, so for gLV we consider the conditional probabilities $\pi_i(s|\underline N_i,\{\underline{N}_j\}_{j \in \partial i})$. 
We express this probability in terms of a product over neighbors of conditional probabilities $\pi^{j\to i}(s| \underline N_j,\underline N_i)$ that a neighbor $j$ of the site $i$ belongs to a sub-extensive cluster of size $s$ in the absence of the edge $(ij)$. 
Our theory then focuses on the probability generating function defined as $H^{j\to i}_{\underline N_i,\underline N_j}(z) = \sum_{s=0}^\infty \pi^{j\to i}(s| \underline N_j,\underline N_i) z^s$ associated to these conditional probabilities. We proceed by finding a self-consistent recursion that allows us to efficiently compute these generating functions. The key assumption we make, and which allows this formalism to be developed, is that the probability distribution that induces the percolation admits a locally tree-like factor graph representation, i.e., it is amenable to study through BP. This is the case for the gLV model under consideration with the BP equation given by Eq.~\eqref{uniform_messages}. In the case of $d$-regular graphs with uniform interactions (see Appendix~\ref{Percolation_Exact} for the general case), the conditional generating functions can be taken uniform (under the replica symmetric assumption). We denote it by $H_{\underline{N},\underline{M}}(z)$, where $\underline{N}$ and $\underline{M}$ are the trajectories of the corresponding site and of one of its neighbors. We show in Appendix~\ref{Percolation_Exact} that  $H_{\underline{N},\underline{M}}(z)$ respects the following recursion
\begin{equation}\label{recursion_MAIN}
\begin{aligned}
H_{\underline{N},\underline{M}}(z)
&=
\indi[N^{t=p+c}=0]
+
z\,\indi[N^{t=p+c}>0]
 \\
&\hspace{-0.5cm}\times \sum_{\{\underline{M}_\ell\}_{\ell=1}^{d-1}}
P\!\left(
\{\underline{M}_\ell\}_{\ell=1}^{d-1}
\,\middle|\,
\underline{N},\underline{M}
\right) 
\prod_{\ell=1}^{d-1}
H_{\underline{M}_\ell,\underline{N}}(z).
\end{aligned}
\end{equation}

The probability $P\!\left(\{\underline M_\ell\}_{\ell=1}^{d-1}\middle| \underline N,\underline M\right)$ denotes the conditional probability that $d-1$ neighbors of a site follow the trajectories $\{\underline M_\ell\}_{\ell=1}^{d-1}$, given that the trajectory of the site is $\underline N$ and that the trajectory of one of its neighbors is $\underline M$. In the context of gLV we obtain  $P\!\left(\{\underline M_\ell\}_{\ell=1}^{d-1}\middle| \underline N,\underline M\right)$  through BDCM
and find that 
\begin{equation}
    P\!\left(\{\underline M_\ell\}_{\ell=1}^{d-1}\middle| \underline N,\underline M\right)\!\!= \!\!\frac{\mathcal{A}\left(\underline{N},\underline{M}\cup \{\underline M_\ell\}_{\ell=1}^{d-1}\right)\prod_{\ell=1}^{d-1}\chi^{\rightarrow}_{\underline{M}_\ell,\underline{N}}}{Z^{\rightarrow} \chi_{\underline{N},\underline{M}}^{\rightarrow}}.\nonumber
\end{equation}
We remark that the recursion in Eq.~\eqref{recursion_MAIN} is rather general and can be applied to any dynamics of the form in Eq.~\eqref{longequation} for which one can determine $P\!\left(\{\underline M_\ell\}_{\ell=1}^{d-1}\middle| \underline N,\underline M\right)$. In order to compute $\phi_{LC}$ we notice that $H_{\underline{N},\underline{M}}(1)$ gives the conditional probability that a site belongs to a sub-extensive cluster and thus one can express $\phi_{ LC}$ via $1-H_{\underline{N},\underline{M}}(1)$ (assuming that the largest connected component is unique). The precise expression for $\phi_{LC}$, derived in Appendix~\ref{Percolation_Exact}, is the following
\begin{widetext}
 \begin{align}\label{phi_LC_main}
     \phi_{LC}
     &=
    \frac{\sum_{\underline{N},\underline{M}}{\indi\left[N^{p+c}> 0\right] \chi_{\underline{N},\underline{M}}\chi_{\underline{M},\underline{N}}}}{\sum_{\underline{N},\underline{M}}{\chi_{\underline{N},\underline{M}}\chi_{\underline{M},\underline{N}}}}
     -\frac{1}{Z_{\rm fac}}\!\!\sum_{\underline N,\{\underline M_\ell\}_{\ell=1}^{d}}\!\!\!\!\!\!\!
    \indi\!\left[N^{t=p+c}>0\right]
     \mathcal A\left(\underline N,\{\underline M_\ell\}_{\ell=1}^{d}\right)
    \prod_{\ell=1}^{d}
    H_{\underline M_\ell,\underline N}(1)\chi_{\underline M_\ell,\underline N}
  .
   \end{align}
   \end{widetext}

In order to compute $\phi_{LC}$, we firstly find the messages $\chi$ with Eq.~\eqref{uniform_messages}, solve the recursion for $H_{\underline{N},\underline{M}}(1)$ in Eq.~\eqref{recursion_MAIN}, and thus determine the $\phi_{LC}$ fraction from Eq.~\eqref{phi_LC_main}. It is interesting to point out that this equation reduces to ordinary site percolation (see~\cite{Site_percolation}), after noticing that in site percolation the state variables  are binary  and the probability $P(\{w_{\ell}\}_{\ell=1}^d| x,y)$ (i.e the corresponding object of $P\!\left(\{\underline M_\ell\}_{\ell=1}^{d-1}\middle| \underline N,\underline M\right)$ in the binary state case) is just \[P(\{w_{\ell}\}_{\ell=1}^d| x,y)=\prod_{\ell=1}^{d-1} \Bigl(\delta(w_\ell=1)p+(1-p)\delta(w_{\ell}=0)\Bigr),\] with $p$ the (site independent) occupation probability (see Appendix~\ref{Percolation_Exact} for details). Also, with another choice of state variable and of $P(\{w_{\ell}\}_{\ell=1}^d| x,y)$, one can also recover the edge percolation equations (see \cite{PercolationBP}).

\subsection{Analysis of dominant attractors} 
In principle, in order to have an exact description of the properties of dominant attractors (and of the dynamical trajectories leading to them), one would need to consider transients of length $p \to \infty$ as the size of the system $S \to \infty$, as in this limit Eq.~\eqref{longequation} converges to the flat measure on all initial conditions (and their associated dynamical trajectories).
In practice, the solution of the BDCM equations at large $p$ is computationally costly, with complexity scaling exponentially in $p$, thus limiting our approach to $p \lesssim 4$.
Nevertheless, if the total free entropy $\Phi_{(p/c)}$ converges rapidly to its $p\to\infty$ value of $\log(K)$ (there are $K^S$ possible initial conditions), then small values of $p$ may be sufficient to cover a significant fraction of the entropy of the basins of attraction.
The associated average values of $\rho_0$, $\eta_l$, and $\phi_{LC}$  would then be expected to approximately describe the typical properties of the system even at moderate $p$. 
This rapid convergence of the entropy to its maximum has been observed in other cellular automata systems \cite{BDCM,CellAuto}, where it is a consequence of the fast relaxation of the dynamics for typical initial conditions. We will see that the rapid convergence of the entropy also holds for the discrete gLV dynamics.

\begin{figure*}[!t]
\centering

\hfill
\begin{minipage}{0.66\textwidth}
    \centering
    \includegraphics{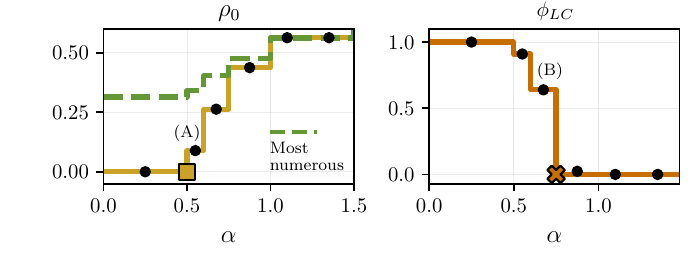} 
\end{minipage}
\hfill
\begin{minipage}{0.32\textwidth}
    \vspace{-1cm}
    \centering
    \begin{tabular}{@{\hspace{4pt}}c@{\hspace{10pt}}c@{\hspace{10pt}}c@{\hspace{4pt}}}
        \hline
        \multicolumn{3}{c}{
        \begin{tabular}{c}
        BDCM entropy $\Phi_{(p/c)}/\log K$ \\
        for $\alpha$ near critical thresholds
        \end{tabular}
        } \\
        \hline
        $p$ & $\alpha=0.55$ (A) & $\alpha=0.675$ (B) \\
        \hline
        $1$ & $0.9650$ & $0.9438$ \\
        $2$ & $0.9999$ & $0.9976$ \\
        $3$ & $0.9999$ & $0.9999$ \\
        $4$ & $0.9999$ & $0.9999$ \\
        \hline
    \end{tabular}
\end{minipage}
\hfill

\vspace{0.4cm}

\begin{minipage}{\textwidth}
    \centering
    \includegraphics{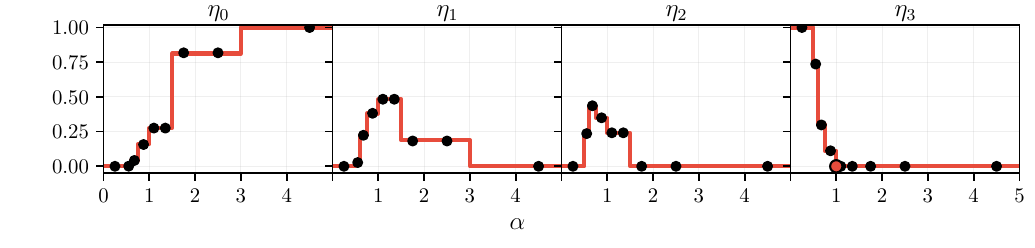}
\end{minipage}

\caption{\textbf{BDCM accurately captures typical dynamics and structure}, illustrated for $K=2$ on $d=3$--random regular graphs.
\textit{(Top)} Extinction fraction $\rho_0$ and largest cluster fraction $\phi_{LC}$ as a function of $\alpha$, computed from BDCM on the dominant $(p/c)$--attractors ($p=3$, $c=2$). Black dots denote numerical simulations ($S=10^4$, averaging 5 simulations, error bars are negligible), lines represent theoretical predictions, showing perfect agreement. Colored markers indicate the extinction (square) and percolation (cross) phase transitions (same notation as in Figure~\ref{fig1}). The table reports the normalized entropy $\Phi/\log(K)$ as a function of $p$ (for $c=2$), showing rapid convergence to the maximal value. 
\textit{(Bottom)} Structure functions $\eta_l$ versus $\alpha$ for the same attractors. The red marker identifies the linear community threshold, where $\eta_3$ vanishes. The same phenomenology is present for other $d$, see Figure~\ref{all_d_K_2} in Appendix~\ref{Additonal_BDCM}.
} 

\label{fig2}
\end{figure*}

\textbf{Summary of the procedure.} 
This summarizes the pipeline required to derive the phase diagram in Figure~\ref{fig1}: given $\alpha, d, K$, one solves numerically the BDCM equations for small transient length $p$ and $c=1,2$ (other possibilities are excluded for this dynamics, see Appendix~\ref{App_proof}), checks that the associated entropy converges fast enough to $\log(K)$, computes the observables $\rho_0, \eta_l$, and $\phi_{LC}$ and identifies the system's phases as described in Section \ref{sec:model}.
We showcase this program for the representative case $K=2$, $d=3$ in Figure~\ref{fig2} (in which we detail the values of the observables as a function of $\alpha$). The phase diagram in Figure~\ref{fig1} has been obtained by also applying this methodology to multiple $d$.
As an additional result, in Appendix~\ref{high_d_limit} we discuss how the system's properties behave for a large degree $d$, leveraging analytical simplifications in the BDCM equations in that limit.

\vspace{-5mm}

\section{Discussion of results}

\vspace{-2mm}

We focus our analysis on the representative case $K=2$, $d=3$, capturing most of the phenomenology, and discuss what changes for larger values of $K, d$ when needed.

\textbf{The dynamics converges rapidly.}
For the theoretical framework to be applicable, the entropy should be close to $\log(K)$ already for small values of $p$. In the table in Figure~\ref{fig2}, we show that this is the case for $K=2$ by reporting the entropy at the two values of $\alpha$ for which convergence to $\log(K)$ is slowest. The fact that $\Phi$ is already so close to $\log(K)$ at $p=3$ is a clear indication of fast dynamical convergence, and validates the correctness of our theoretical analysis.

The same fast convergence to $\log(K)$ is observed also for higher $d$. As $K$ is increased, convergence to the maximal entropy $\log(K)$, however, becomes slower. Nevertheless, for $K=3$, the largest computationally accessible value, $p=4$, already captures about $99\%$ of the entropy for most values of $\alpha$, and remains above $95\%$ even near the percolation threshold, where convergence in entropy is slowest.

We additionally remark that it is particularly striking that the discrete gLV model with small $K$, despite its fast relaxation, reproduces the same phenomenology as both the continuous model and the higher-$K$ cases. In particular, Figure~\ref{comparison_d_3} in Appendix~\ref{Appendix_Numeric} shows that the qualitative behavior of $\rho_0$ and $\phi_{LC}$ at convergence is similar across several orders of magnitude in $K$, and agrees with the continuous model in Eq.~\eqref{continuous_glv}.

\textbf{Dominant attractors are 2-cycles.} 
To compute the values of the observables, we must first determine whether the dominant attractors are fixed points $c=1$ or 2-cycles $c=2$. To this end, we compute the entropy $\Phi_{(p/c)}$ for $c=1,2$ for multiple values of $p$ and $\alpha$. We find that, unless $\alpha$ is extremely large, the dominant attractors are cycles of length $c=2$, as their entropy is the largest for every $p$. This implies that exponentially more initial conditions converge to 2-cycles compared to fixed points (see Figure~\ref{c1vsc2} in Appendix~\ref{cyclefixedpoint}).  
For very large $\alpha$, on the other hand, the dominant attractor is a (trivial) independent set fixed point: each occupied site is isolated (for example, in the case $K=2$, $d=3$ we have $\eta_0=1.0$ for $\alpha\geq 3.0$). 

The same behavior is observed, both from BDCM and through empirical simulations, for each value of $K$ and $d$ considered (see Table~\ref{Diagram_K=3} for a list of tested $(K,d)$ parameter pairs).

\textbf{Theory matches with numerical experiments.} 
We show in Figure~\ref{fig2} that for the dominant attractor, already $p=4$ is sufficient to capture the typical behavior of the system: the values of $\rho_0, \phi_{LC}$ and $\eta_l$ match perfectly with those obtained from numerical simulations over the full range of  $\alpha$. The fact that $\phi_{LC}$ matches the numerical simulation also confirms the assumption we made about the uniqueness of the largest connected component. 
We report the same comparison for $K=2$ and different $d$ in Figure~\ref{all_d_K_2} in Appendix~\ref{Additonal_BDCM}, to show that the same good matching between theory and simulations occurs. 
We report the values of $\rho_0$, $\phi_{LC}$ and $\eta_l$ for $K=3$ and $d=3$ in Figure~\ref{appendix_K_3} in Appendix \ref{Additonal_BDCM}, and observe that the phenomenology is largely compatible to that of the $K=2$ case. 

\textbf{The most numerous attractors are not the dominant ones.} We now explicitly show that taking into account transients is crucial to predict the behavior of the system at convergence, as the most numerous attractors have different properties than the dominant ones. 
To study the most numerous attractors, it is sufficient to consider the measure in Eq.~\eqref{longequation} with $p=0$ and without the term $\prod_{i=1}^S \mathbbm{1}[N_i^{t=1} > 0]$ (this describes the flat measure over $c$-cycles). What we find is that the $\rho_0$ fraction computed for the most numerous cycles (again, with $c=2$) is positive even at $\alpha < \alpha_{\rm ext}$, meaning that in the fully occupied phase the most numerous attractors have a finite fraction of vacant sites even though the dynamics converges to fully-occupied attractors (see Figure~\ref{fig2} top-left).
This implies that the randomly initialized dynamics does not converge to the most numerous attractor.
The same phenomenology is observed at other values of $d$ and $K$.

\begin{figure*}[!t]
\begin{minipage}[t]{0.8\textwidth}
    \centering
    \vspace{0pt}
    \includegraphics{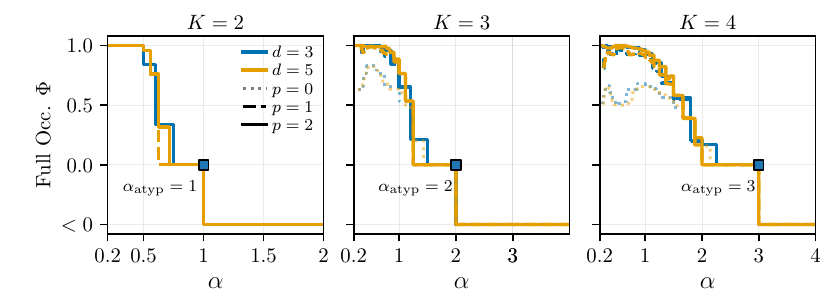}
\end{minipage}
\hfill
\begin{minipage}[t]{0.19\textwidth}
\vspace{0.5cm}
    \centering
    \scriptsize
    \setlength{\tabcolsep}{4pt}
    \renewcommand{\arraystretch}{1.4}
    \begin{tabular}{ccc}
        \hline
        $K$ & $\alpha_{\rm atyp}$ & $\alpha_{\rm ext}$ \\
        \hline
        \multicolumn{3}{c}{$d\leq 6$} \\
        \hline
        $2$ & $1$ & $1/2$ \\
        $3$ & $2$ & $2/3$ \\
        $4$ & $3$ & $3/4$ \\
        \hline
        \multicolumn{3}{c}{$d\to\infty$} \\
        \hline
        $2$ & $\approx 1.0$ & $\approx 0.66$ \\
        $3$ & $\approx 1.33$ & $\approx 1.0$ \\
        $4$ & $\approx 1.48$ & $\approx 1.1$ \\
        \hline
    \end{tabular}
\end{minipage}

\caption{
\textbf{Full occupied entropy and atypical full-survival threshold.}
\textit{(Left)} The entropy $\Phi^{\rm Full.Occupied}_{(p/c)}$, which counts the log-number of atypical initial conditions leading to a fully occupied attractor, for multiple values of $K$, $d$, and $p$. For $K=2$, $d=3$, the entropies coincide for all $p$. For visual clarity all the negative entropies correspond to the marker $<0$.
\textit{(Right)} Threshold values $\alpha_{\rm atyp}$ and $\alpha_{\rm ext}$ for finite $d$ and in the limit $d\to\infty$. We refer to Figure~\ref{figure_d_infty_atyp} in Appendix~\ref{High_d_appendix} for the entropies in the limit $d\to\infty$, where we also report the critical threshold.
}
\label{finalfigureB}
\end{figure*}

\textbf{Subdominant fully occupied attractors.}
We now investigate whether there exist fully occupied subdominant attractors outside of the fully occupied phase. Indeed, even at $\alpha> \alpha_{\rm ext}$ there may exist exponentially rare initial conditions that lead to a fully occupied community. 
To check the existence of such subdominant attractors, we consider the probability distribution in Eq.~\eqref{longequation} with an added indicator function that prevents sites from ever becoming extinct ($\indi[N_i^{t=p+c}>0]$), and apply the BDCM method again. 
We consider $c=2$, since for almost every value of the interaction strength
$\alpha$ (i.e., up to a set of zero measure), all the fixed points of the dynamic ($c=1$) necessarily present extinction (see the final remark in Appendix~\ref{App_proof}). 
Regarding the transient length $p$, we take $p=0$ to probe the existence of the subdominant attractor (existence does not depend on the transient), and $p>0$ to probe the size of the basin of attraction.  
The associated entropy density $\Phi^{\rm Full. Occupied}_{(p/c)}$ gives the log-number of atypical initial conditions leading to a fully occupied community, so when it is non-negative it means that there exist initial conditions (provided the correctness of the RS assumption) leading to a fully occupied community. We then label $\alpha_{\rm atyp}\geq \alpha_{\rm ext}$ as the largest interaction strength with a non-negative $\Phi^{\rm Full. Occupied}_{(p/c)}$ entropy. When $\alpha_{\rm atyp}>\alpha_{\rm ext}$, then it means that there is a region of interaction strengths for which typical initial conditions will converge to an attractor that presents extinction, while there may exist atypical initializations leading to a fully occupied community.

In Figure~\ref{finalfigureB} we show, for $K\in\{2,3,4\}$, $d\in\{3,5\}$ and multiple values of $p$, the entropy $\Phi^{\rm Full. Occupied}_{(p/c)}$ as a function of $\alpha$ (see Figure~\ref{figure_d_infty_atyp} in Appendix~\ref{High_d_appendix} for the case $d\to\infty$), and report the values of $\alpha_{\rm atyp}$ and $\alpha_{\rm ext}$. We observe that the gap $\alpha_{\rm atyp}-\alpha_{\rm ext}$ is always strictly greater than zero, it increases with $K$, and it persists up to $d\to \infty$. This shows that the existence of rare initial conditions that lead the system to be fully occupied for a range $\alpha_{\rm ext}\leq \alpha\leq \alpha_{\rm atyp}$ is a robust phenomenon. 
We also remark that as $K$ and $d$ increase, the size of the basins of attraction of these subdominant attractors increases too. This can be seen from the entropy plots in Figure~\ref{finalfigureB}, where we show that the entropy  $\Phi^{\rm Full. Occupied}_{(p/c)}$ for $K=3,4$ at $p=0$ is much lower than the entropy at $p>0$, signifying that at higher $K$ it is not necessary to initialize the system directly in these attractors in order to reach them (the opposite is true for $K=2$, for which instead the entropy at $p=0$ coincides almost everywhere with the entropy at higher $p$).

\vspace{-5mm}

\section{Conclusion}
\vspace{-2mm}
In this work, we analyzed a discrete version of the gLV model, in which competitive interactions between sites are spatially structured. We developed a dynamical theory of this model, rooted in the backtracking dynamical cavity method, which allowed us to show that depending on the interaction strength, the dynamics will converge to different types of dynamical attractors. As competition increases, the community transitions from a fully occupied phase to regimes with extinction and fragmentation in spatially separated clusters, and we show that the topology of said clusters is also determined by interactions. 
The first of our main contributions is the theoretical determination of sharp thresholds between these phases. This gives an analytical description of phenomenology observed in real ecosystems \cite{Kefi2,Review, ge2023hidden, von2001diversity,rietkerk2002self} and reproduced in qualitatively similar computational models \cite{marcus2022local, vega2011effects, Kefi1}, mostly studied numerically. 

The second of our main contributions is methodological. In order to study percolation at convergence for the discrete gLV model, we developed a new message-passing analysis. This method generalizes previous results on bond-percolation, where each edge is present in the underlying graph independently, to the more challenging setting of dependent edges, capturing, for example, percolation induced by dynamical systems and opening the door to studying these phenomena in a wide variety of settings. In particular, we foresee applications of this method  not only in ecosystem ecology, but also, for instance, in social sciences, epidemic spreading, and cellular automata modeling, fields that are naturally concerned with percolation phenomena \cite{xie2022indirect, Kefi2, newman2002spread}.

The dynamical analysis additionally reveals that the most numerous attractors are not the ones toward which a randomly initialized trajectory will converge, showing the importance of dynamical descriptions when studying these systems. This complements other analytical works on high dimensional ecological systems, such as the Lotka Volterra model, which mainly focus on the structure and linear stability of the equilibria of the dynamics  \cite{bunin2017ecological,marcus2022local,Kak-Rice,tonolo2026generalized,ValentinaRos,advani2018statistical,cui2020effect}, without taking into account transients. 

Finally, our methods give access to subdominant attractors of the dynamics, which are attractors that are almost never reached from random initializations, yet remain accessible if the system is started from carefully chosen initial conditions. In particular, we show that for certain interaction strengths, there exist subdominant attractors characterized by a fully occupied community, even though random initial conditions lead to vacant sites. This result  provides a theoretical perspective on observations in restoration ecology (which studies under which conditions an ecological system can be prevented  from presenting extinction \cite{silliman2024harnessing}). In particular, it has been observed that seed dispersal and seeding density \cite{orth2020restoration,wells2024seeding}, together with precise spatial arrangements \cite{corbin2012applied,grafnings2023spatial,kjaer2024spatial} (these, in our framework, correspond to initial conditions) can strongly influence the final state of grassland, seagrass and forests. Carefully designed initial configurations (as done for instance in \cite{jankola2026minority}) can therefore play a decisive role in determining whether an ecological community persists or collapses.

\vspace{-8mm}
\section{Acknowledgment}
\vspace{-4mm}
We acknowledge Jacopo Grilli, Onofrio Mazzarisi, Chiara Cammarota and Giulio Biroli for insightful discussions.  

\vspace{-6mm}
\section*{Data availability}
\vspace{-4mm}
The code and data needed to reproduce the analyses presented can be found on \href{https://github.com/SPOC-group/Clustering-and-percolation-in-ecological-communities-A-dynamical-theory}{GitHub}.

\bibliography{references/references}

\clearpage
\onecolumngrid
\appendix

\section{Numerical analysis of the discrete gLV model for different topologies, higher degrees $d$, and diverse interaction patterns: the phase transition phenomenology of the main text is robust}\label{Appendix_Numeric}
\subsection{Discrete gLV on a grid.} 
\begin{figure}[!h]
    \centering
    \includegraphics{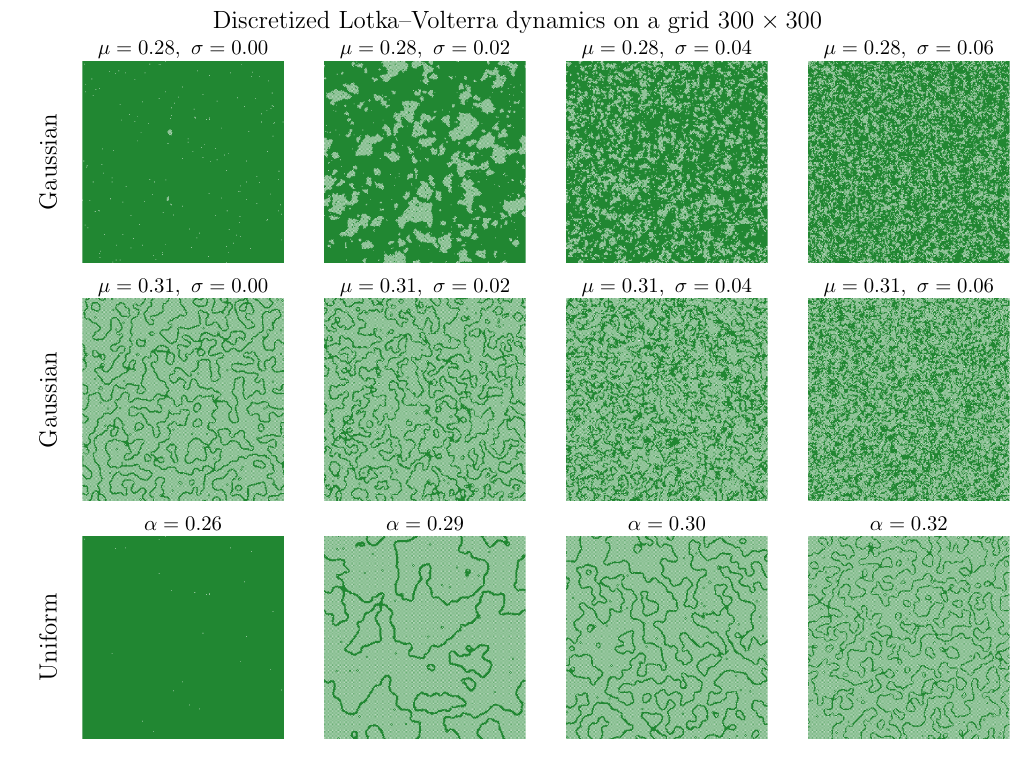}
    \caption{\textbf{Simulation of discrete gLV model on a grid, at convergence.} (\textit{First two rows)} Coupling of neighboring sites sampled from a truncated Gaussian, with standard deviation $\sigma$ and mean $\mu$. The white spots indicate extinct sites (at convergence), dark green occupied sites. The fragmentation of the community is visible in the first row. (\textit{Third row}) Uniform coupling case, where the emergence of ring-like structures can be observed. Both simulations are done for $K=15$.}
    \label{fig:grid_lotka_volterra}
\end{figure}

We show in Figure~\ref{fig:grid_lotka_volterra} that when the system is simulated on a grid (each site has 4 neighbors, right, left, top and bottom) more complex spatial structures can emerge. In particular we see the formation of ring-like patterns of occupied sites, which is consistent with what is observed in real ecosystems \cite{ge2023hidden}, and we graphically observe the formation of spatially clustered communities (see the fragmentation/patchiness phenomena discussed in \cite{Kefi2, ManorStructured,vega2011effects}). This highlights the fact that our model captures the main behavior of real ecosystems. We show numerical simulations both in the case of a uniform interaction strength $\alpha$ (in order to consider the case of an isotropic environment), and also the case of interactions sampled from a truncated Gaussian ( $\alpha_{ij}>0$, so the system is strictly competitive) with finite mean and finite variance. In the case of samples from the truncated Gaussian we call $\mu$ the mean and $\sigma$ the variance. 

\subsection{The discrete dynamic in the limit of large $K$.} 
\begin{figure*}[!t]
    \centering
    \begin{minipage}{\textwidth}
        \centering
        \includegraphics{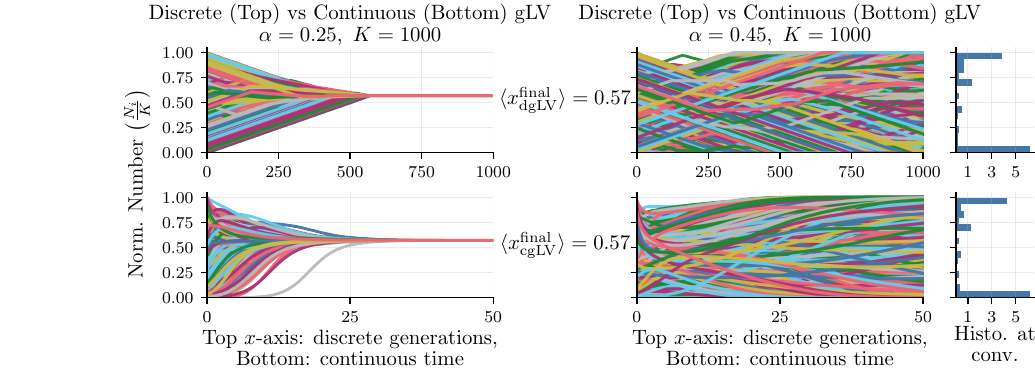}
    \end{minipage}
    \vspace{0.1mm}
    \begin{minipage}{\textwidth}
        \centering
        \includegraphics{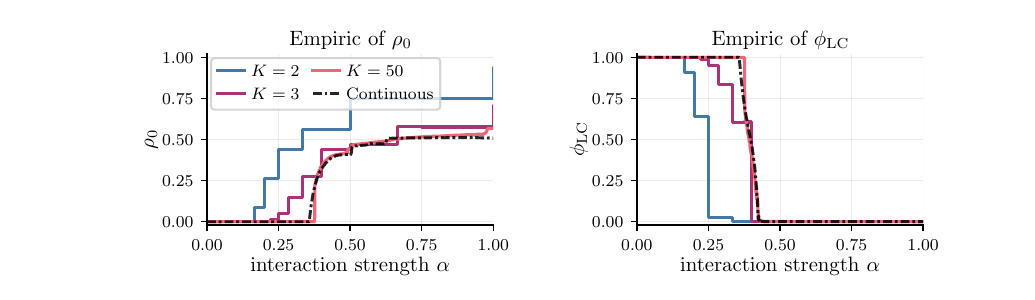}
    \end{minipage}
    \hspace{0.01\textwidth}
   
\caption{\textbf{Empirical comparison of discrete generalized Lotka-Volterra with its continuous counterpart, uniform interactions.} Simulations on a random $3-$regular graph with $S=10^5$ show that discrete gLV closely reproduces the behavior of continuous gLV. \textit{(Top row)} For small $\alpha$ both models reach a uniform state, whereas for larger $\alpha$ extinction and heterogeneous biomass levels emerge. This is consistent with what is reported in the usual literature of continuous gLV model, on random regular graphs (see for instance \cite{marcus2022local, Fabian, tonolo2026generalized}). The histograms of the normalized biomass at convergence, shown alongside the trajectories, are nearly identical in the two models, indicating that not only the mean behavior but the full distribution is well captured by discrete gLV, when $K$ is large. \textit{(Bottom row)} The observables $\rho_0$ (fraction of unoccupied/extinct sites) and $\phi_{LC}$ (fraction of sites in largest connected component) show good agreement already for moderate $K$ for all investigated values of $\alpha$, and remain qualitatively similar even for smaller carrying capacities.}
\label{comparison_d_3}
\end{figure*}

We show that for large $K$ the discrete gLV model in Eq.~\eqref{discrete_glv} approximates very well the continuous model in Eq.~\eqref{continuous_glv}. We carry out  numerical integration of the differential equations for continuous dynamical rule in Eq.~\eqref{continuous_glv} for systems of size $S\ge 10^4$ (the simulations for the discrete case are also carried out for a system of $10^4$ sites). We do this comparison in the case of $d=3$ with uniform interactions (Figure~\ref{comparison_d_3}), where we show the shape of the trajectory for the continuous and discrete case over time (we plot the ``normalized" biomasses $N_i/K$). We also show the extinction fraction $\rho_0$ and fraction of sites in the largest connected component $\phi_{LC}$ as a function of the interaction strength. This shows that the discrete version of the gLV model in Eq.~\eqref{discrete_glv} approaches (at least on a qualitative level) the behavior of the continuous one which is usually studied in the literature (see \cite{tonolo2026generalized,marcus2022local}) as $K$, the carrying capacity, is increased. As in the main text, we scale the $\alpha$ factor by $d$ (i.e. the growth rate in Eq.~\eqref{growthrate} presents the standard interaction strength $\alpha/d$.)
\subsection{Numerical simulations on tree-like graphs with non-uniform interactions.} 
\begin{figure}
    \centering
    \includegraphics[width=\textwidth]{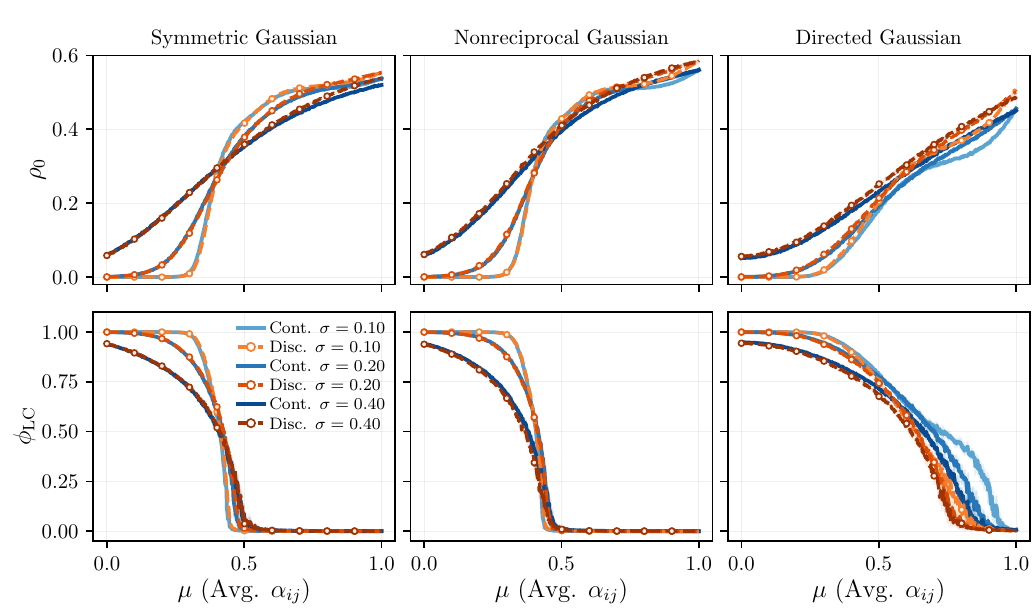}
    \caption{\textbf{Numerical results for heterogeneous interactions.} Numerical measurements of the occupied fraction $\rho_0$ and of the largest-cluster fraction $\phi_{LC}$ for the discrete model on random $3$-regular graphs with $S=10^4$ sites and carrying capacity $K=100$, for symmetric, nonreciprocal, and directed truncated Gaussian couplings $\alpha_{ij}$. For comparison, we also show the corresponding results for the continuous gLV model. As the mean interaction strength $\mu$ increases, $\phi_{LC}$ decreases to zero while $\rho_0$ increases, with only quantitative differences between the different coupling ensembles, showing that fragmentation and percolation are robust to interaction heterogeneity.}
    \label{DifferentCouplings}
\end{figure}

We simulate our dynamics for heterogeneous interaction strengths and measure the observables $\rho_0$ and $\phi_{LC}$, the results are shown in Figure~\ref{DifferentCouplings}. In particular, we consider the following distributions for the couplings $\alpha_{ij}$. For the discrete model we fix the carrying capacity to $K=100$, and the simulations are done for a system with $S=10^4$. The topology of the network is still a $3$-regular graph.
\begin{itemize}
    \item[\textit{1)}] \textit{Symmetric Gaussian.} $\alpha_{ij}$ sampled from a truncated Gaussian distribution with the constraint $\alpha_{ij}=\alpha_{ji}$ (symmetric interactions). We consider distinct values of the variance $\sigma$, while we call $\mu$ the average interaction strength
     \item[\textit{2)}] \textit{Nonreciprocal Gaussian.} $\alpha_{ij}$ sampled from a truncated Gaussian distribution without any symmetry constraint (asymmetric interactions).
     \item[\textit{3)}] \textit{Directed Gaussian.} $\alpha_{ij}$ sampled from a truncated Gaussian distribution with the constraint that if $\alpha_{ij}>0$ then $\alpha_{ji}=0$, corresponding to a directed network (i.e., interactions are one-way).
\end{itemize}
For comparison, we also simulate the continuous gLV model on systems of the same size. We find that the behavior of $\rho_0$ and $\phi_{LC}$ is qualitatively similar across all sampling schemes and (average) interaction strengths (with the most significant difference observed for the directed Gaussian case), and essentially identical at small interaction strengths. Moreover, as the average interaction strength increases, $\phi_{LC}$ decreases to zero while $\rho_0$ increases, in agreement with the behavior observed in the uniform case. This shows that the fragmentation and percolation phenomena are not artifacts of uniform interactions, but are robust to heterogeneity in the interaction structure.
\section{The relationship between the consumer resource model and the gLV model}\label{Appendix_CR}
We review the relationship between the consumer resource model and the gLV model, in order to highlight that the gLV model can be seen as a limiting case of the consumer resource model, and that the interaction strengths are related to environmental (tunable) conditions. This type of equivalence is well known, as it has already been shown  in the pioneering work of MacArthur \cite{MacArthur,MacArthurLotkaVolterraCR}. Nevertheless, we find it useful to remark this relationship, also to highlight that it is preserved in our specific case, where we consider a (spatial) discrete version of gLV.  We start from the equations of the consumer resource (CR) model, where we denote by $\{R_\mu\}_{\mu=1}^{M}$ the resources ($M$ total number of resources), and with $\{N_i\}_{i=1}^S$  the sites (or species). We want to show how to recover gLV, both the discrete version in Eq.~\eqref{discrete_glv} and the continuous one in Eq.~\eqref{continuous_glv}, from the CR model. We start by assuming that the biomasses in each site can access nourishment only from a subset of resources, which are accessible (due to spatial proximity) to it. The consumer resource equations are as follows (for simplicity, we use the version of \cite{Fant,MacArthur,MacArthurLotkaVolterraCR})
\begin{align}
    \frac{dR_{\mu}}{dt}&=R_{\mu}\Bigr [K_\mu-R_{\mu} -\sum_{j} a_{j\mu}N_j\Bigl ]\label{B1}\\
    N_i(t')&=F\Bigl(N_i(s), \sum_{\mu}R_{\mu}(s)b_{\mu i}\Bigr| s\leq t'\Bigl),\label{eqB2}
\end{align}
where $a_{j\mu}>0$ tells if site/species $i$ consumes resource $\mu$, while $b_{\mu i}$ determines if site $i$ obtains a benefit from consuming resource $\mu$. $F(\cdot)$ is a general function that describes the dynamics of the biomass $N_i$, and can be chosen freely. In particular in this work we will consider two possibilities. If one considers the usual interpretation of the CR model (as in \cite{Fant,MacArthur,MacArthurLotkaVolterraCR}) then $F$ is chosen as
\begin{equation}\label{B4}
        N_i(t')=N_i^{t=0}+ \int_{0}^{t'}\, ds\, N_i(s)\Bigl[K_i-N_i(s)+\sum_{\mu}R_{\mu}(s)b_{\mu i}\Bigr].
\end{equation}
Indeed, one can note that this is just the integrated representation of a differential equation (i.e. the ODE system composed of Eq.~\eqref{B1} and the derivative of  Eq.~\eqref{B4} is the ODE system of the CR model).

A second choice of $F$, which is the one more related to this work (see Eq.~\ref{continuous_glv}), is given by

\begin{equation}
        N_i^t=N_i^{t-1}+ {\rm sgn} \Bigl[N_i^{t-1}\Bigl(K_i-N_i^{t-1}+\sum_{\mu}R_{\mu}^{t}b_{\mu i}\Bigr)\Bigr].
\end{equation}
The CR equations describe the coupled dynamics of consumers and resources: consumer biomasses $N_i$ grow by consuming resources $R_\mu$, while resource abundances are depleted through this consumption.

Then, under the standard approximation that the resources equilibrate very quickly compared to the $N_i$, i.e. that there is some separation of the timescales for the dynamic of consumers and resources, the average value of each $R_{\mu}^*$ at equilibrium is
\begin{equation}
R_{\mu}^*\approx K_{\mu}-\sum_{j} a_{j\mu}N_j(s).
\end{equation}
Substituting this expression in equation~\eqref{eqB2} gives
\begin{equation}
    N_i(t')=F\Bigl(N_i(s),  \sum_{\mu}K_{\mu}b_{\mu i}-\sum_{j,\mu} a_{j\mu}N_j(s)b_{\mu i}\Bigr| s\leq t'\Bigl).
\end{equation}
This means that, after defining $\alpha_{ij}=\sum_{\mu}a_{j\mu}b_{\mu i}$, the evolution of the $N_i$ is given by either a continuous or discrete gLV model with interaction strength $\alpha_{ij}$, depending on which $F(\cdot)$ has been considered : \begin{align}
     N_i(t')&=N_i^{t=0}+ \int_{0}^{t'}\, ds\, N_i(s)\Bigl[K_i-\sum_{\mu} K_{\mu}b_{\mu i}-N_i(s)-\sum_{j}N_{j}(s)\alpha_{ij}\Bigr]  \quad \text{Continuous gLV model, as in Eq.~\eqref{continuous_glv}},\label{eqB7} \\ 
     N_i^{t}&=N_i^{t-1}+ {\rm sgn} \Bigl[N_i^{t-1}\Bigl(K_i-\sum_{\mu} K_{\mu}b_{\mu i}-N_i^{t-1}-\sum_{j}N_j^{t}\alpha_{ij}\Bigr)\Bigr] \quad \text{discrete gLV model, as in Eq.~\eqref{discrete_glv}}.  \label{eqB8}
\end{align}
$\alpha_{ij}$ encodes the interaction topology. Indeed two sites (or specie) interact only if $\alpha_{ij}\neq 0$. This happens if there is at least a resource for which the two sites compete for, i.e. a resource $\mu^*$ from which the biomass from site $i$ takes advantage of ($a_{\mu^*i}>0$), and a resource consumed also by the biomass at site $j$ ($b_{j\mu^*}>0$). In the spatial and uniform  interpretation of gLV one can assume that all sites are equivalent to each other, and that they consume resources that are spatially close to them. Then $\alpha_{ij}=\alpha$ if $i$ and $j$ are neighboring, while $\alpha_{ij}=0$ if they are not neighboring. The relation of the interaction strength with environmental conditions follows from the fact that both $a_{\mu i}$ and $b_{\mu i}$ are in principle related to the environment. Indeed, pH, precipitation and heat can, among other factors, alter both coefficients. This is verified experimentally in \cite{gore2025transition}. For analytical tractability, in our case, we assumed that the carrying capacities are uniform among the sites (or species). Indeed the $\kappa$ and $K$ carrying capacity that appear in Eq.~\eqref{continuous_glv}  and Eq.~\eqref{discrete_glv} can be identified with the term $K_i-\sum_{\mu}K_{\mu}b_{\mu i}$ that appears both in equation Eq.~\eqref{eqB7} and Eq.~\eqref{eqB8} (with the understanding that $K_i-\sum_{\mu}K_{\mu}b_{\mu i}$ represents a continuous carrying capacity in Eq.~\eqref{eqB7}, while it is discrete in Eq.~\eqref{eqB8}), if one assumes that the term $K_i-\sum_{\mu}K_{\mu}b_{\mu i}$ is overall $i$ independent. Furthermore, we assumed the positivity of that term.

\section{Technical remarks about the discrete gLV model and proof that at most cycles of length 2 can exist in the dynamic}\label{App_proof}

\textbf{The discrete generalized Lotka-Volterra Model is a sign dynamic.}
    The discrete gLV model in Eq.~\eqref{discrete_glv} can be written in the following way\begin{equation}
    \label{D1}
    N_i^{t}=N_i^{t-1}+\text{\rm sgn}\Bigl[N_i^{t-1}\bigl(K-\sum_{j=1}^{S}A_{ij}N_j^{t-1}\bigr)\Bigr]
    \end{equation}
    where $\text{sgn}(x)$ is the sign function, with the convention $\text{sgn}(0)=0$, and $A_{ij}>0$ is the interaction matrix with convention $A_{ii}=1$. When the system is uniformly interacting we have $A_{ij}=a_{ij}\alpha$ ($i\neq j$), where  $\{a_{ij}\}$ is the adjacency matrix of the underlying graph. In the case of $d$-regular graph, in this work, we chose to rescale the interaction strength by the degree of the graph, i.e. to take $A_{ij}=a_{ij}(\alpha/d)$ ($i\neq j$). 
    
Note that, given that the interaction matrix has non-negative entries, any site initialized with
$N_i^{0}>K$ eventually enters the region $N_i^t\leq K$. Indeed, whenever
$N_i^t>K$, the growth rate is negative,
\[
N_i^t\left(K-N_i^t-\sum_{j=1}^{S}A_{ij}N_j^t\right)<0,
\]
and it remains negative until $N_i^t\leq K$. Moreover, once all sites satisfy
$N_i^t\leq K$, they remain below $K$ under the dynamics. Let's call $T'$ the time it takes for all the $N_i^t> K$ to go below $K$.

We also define the set $S^*(t)$ of sites $i$ with $N_i^t>0$. Zero is absorbing so $S^{*}(t)\subseteq S^*(t-1) $ for all $  t $. We call $\tilde{T}$ the times it takes to $S^*(t)$ to ``converge" to a limit set $S^*$. For now we consider $S$ finite so that $\tilde T$, $T'$ are also finite due to the discrete states. Then, when taking the limit $S\to \infty$, these times may diverge. In that case, our results must be understood with the specification that one needs to look at the system for ``long" time ($O(S)$) in order to be assured of dynamical convergence. Note that the time it takes to reach an attractor is surely greater than $T^*=\max [T',\tilde T]$.
    
\textbf{Multiple values of $\alpha$ correspond to the same dynamics in a uniformly interacting gLV model on a $d$-regular graph.}
Consider a discrete generalized Lotka-Volterra model with uniform interactions on a $d$-regular graph. In this setting, several values of the interaction strength $\alpha$ can induce exactly the same dynamics. More precisely, the dynamics is unchanged as long as $\alpha$ varies between two consecutive values of the set
\begin{equation} \{\alpha^*_{l}\}_{l=1}^{l_{\rm max}} = \Bigl\{ \frac{d(K-N)}{h} \Bigm| 0\leq N\leq K,\; 1\leq h\leq dK \Bigr\}, \end{equation}
where $l_{max}$ is the cardinality of the set. Indeed, this is precisely the set of values of $\alpha$ for which
\[
K-N-\frac{\alpha}{d}\sum_{j\in\partial i}N_j=0
\]
for some value of  $N$ and some local neighborhood configuration
\[
h=\sum_{j\in\partial i}N_j\leq dK .
\]

To see why the dynamics is constant between two consecutive values of this set, we order the distinct elements of $\{\alpha_l^*\}_{l=1}^{l_{\rm max}}$ increasingly and we consider two values $\alpha,\alpha'\in(\alpha_{l}^*,\alpha_{l+1}^*)$, with  $\alpha_{l}^*,\alpha_{l+1}^*\in\{\alpha^*_{l}\}_{l=1}^{l_{\rm max}}$ (they are subsequent). The update rule depends only on the sign of
\[
N_i^t\left(K-N_i^t-\frac{\alpha}{d}\sum_{j\in\partial i}N_j^t\right).
\]
If $N_i^t=0$, then $N_i^{t+1}=0$, independently of the value of $\alpha$. If instead $N_i^t>0$, and if the  sign of the growth term  between $\alpha$ and $\alpha'$ is different, then there exist some $\tilde\alpha\in(\alpha,\alpha')$ such that
\[
K-N_i^t-\frac{\tilde\alpha}{d}\sum_{j\in\partial i}N_j^t=0 .
\]
But such a value $\tilde\alpha$ would necessarily belong to the set $\{\alpha_l^*\}_{l=1}^{l_{max}}$, contradicting the fact that $\alpha$ and $\alpha'$ lie between two consecutive elements of that set. Therefore, the sign of the growth rate is the same for every site and every configuration, and the induced dynamics is identical throughout the interval.\newline

\textbf{Cycles are at most of length 2.}  Given a discrete gLV model of the type in Eq.~\eqref{D1}, with uniform interactions on a $d$-regular graph, the dynamics admits only fixed points or cycles of length $2$.  More generally, for a symmetric interaction matrix $A_{ij}$, the discrete gLV model admits only fixed points or cycles of length two. We provide here a proof.

We start by defining the following energy function:
\begin{equation}
    E(\vec{N}^t,\vec{N}^{t-1})=-\sum_{i,j\in {S}^{*}}{N_i^tN_{j}^{t-1}A_{ij}}+K\sum_{i\in {S}^{*}}(N_i^{t}+N_i^{t-1}),
\end{equation}
which is defined for $t>T^*$, where $T^*$ and $S^*$ were introduced above. The sum is done only on species with $N_i,N_j> 0$ . Furthermore, given that for  $t>T^*$ we have $0\leq N_i^{t}\leq K$,  the energy will present a lower bound and an upper bound. We now compute the energy difference $\Delta E(t)$ of the dynamic (for now generally assuming $A_{ij}=A_{ji}$):
\begin{equation}
\label{D4}
\begin{split}
    \Delta E(t)&= E(\vec{N}^t,\vec{N}^{t-1})-E(\vec{N}^{t-1},\vec{N}^{t-2})\\
    &=-\sum_{i,j\in {S}^{*}}{N_i^tN_{j}^{t-1}A_{ij}}+\sum_{i,j\in {S}^{*}}{N_i^{t-1}N_{j}^{t-2}A_{ij}} \\ &\quad \, +K\sum_{i\in {S}^{*}}(N_i^{t}+N_i^{t-1})-K\sum_{i\in {S}^{*}}(N_i^{t-1}+N_i^{t-2})\\
    &=\sum_{j\in {S}^{*}} {[N_j^{t-2}-N_{j}^t]}\sum_{i\in {S}^{*}}A_{ij}N_i^{t-1}+K\sum_{j\in {S}^{*}}{[N_j^{t}-N_j^{t-2}]}\\
    &=\sum_{j\in {S}^{*}} {[-N_j^{t-2}+N_{j}^t]}\Bigl\{-\sum_{i\in {S}^{*}}A_{ij}N_i^{t-1}+K\Bigr\},
\end{split}
\end{equation}
where we used the fact that $A$ is symmetric in order to exchange indices. Using the expression of the dynamics, the expression can be simplified further: \begin{equation}
\begin{gathered}
N_i^t=N_i^{t-1}+\text{sgn}\Bigl[N_i^{t-1}\bigl(K-\sum_{j\in S^*}N_j^{t-1}A_{ij}\bigr)\Bigr],\\
\Delta E(t) = \sum_{j\in S^*} {\Bigg\{-N_j^{t-2}+N_j^{t-1}+\text{sgn}\Bigl[N_j^{t-1}\bigl(K-\sum_{i\in S^*}A_{ij}N_i^{t-1}\bigr)\Bigr]\Bigg\}}\Bigl\{-\sum_{i\in S^*}A_{ij}N_i^{t-1}+K\Bigr\}.
\end{gathered}
\end{equation}
Note that we can sum only over $S^*$ when defining the dynamic~\eqref{D1} as we are looking at $t>T^*$ and extinct species do not contribute. Both $i$ and $j$ belong in $S^*$. Thus, we can drop the $N_i^t$ from the sign function, as $N_i^{t-1}> 0$. Secondly, given our dynamic, $\big|N_{j}^{t-2}-N_{j}^{t-1}\big|\leq 1$. Let's now call $g_{j}^{t-1}=\Bigl\{-\sum_{i\in S^*}A_{ij}N_i^{t-1}+K\Bigr\}$. The following then holds
\begin{equation}
\label{D5}
\begin{split}
    \Delta E(t) &=\sum_{j\in S^*} {\Bigg\{-N_j^{t-2}+N_j^{t-1}+\text{sgn}[g_j^{t-1}]\Bigg\}}g_j^{t-1}\\
    &=\sum_{j\in S^*}|g_j^{t-1}|\Bigl\{\text{sgn}[g_j^{t-1}](N_j^{t-1}-N_j^{t-2})+1\Bigr\}\geq 0.\\
\end{split}
\end{equation}
So the energy difference is either $0$ or positive, for all $t>T^*$. Given that the energy is upper-bounded, this means that there exists a time $T\geq T^*$ after which $\Delta E(t)=0$ for each $t\geq T$. In particular, inspecting equation (\ref{D5}), we note that $\Delta E(t)=0$ only when \begin{equation}\label{bigclaim}
\forall j\in S^*\quad \text{either}\quad  N_j^{t-2}=N_{j}^t\, \quad \text{or} \quad N_j^{t-1}=N_{j}^t, 
\end{equation}
Indeed, either $g_j^{t-1}=0$, which happens if $N_j^{t-1}=N_{j}^t$, or $\Bigl\{\text{sgn}[g_j^{t-1}](N_j^{t-1}-N_j^{t-2})+1\Bigr\}=0$.
But we have that \begin{equation}
    \text{sgn}[g_j^{t-1}](N_j^{t-1}-N_j^{t-2})+1=(N_j^{t}-N_j^{t-1})(N_j^{t-1}-N_j^{t-2})+1,
\end{equation}
so $(N_j^{t}-N_j^{t-1})$ and $(N_j^{t-1}-N^{t-2})$ must have opposite signs. However, given the fact that $i,j$ have $N_i,N_j> 0$  we must have either $N_j^{t}=N_{j}^{t-1}-1=N_{j}^{t-2}$ or $N_j^{t}=N_j^{t-1}+1=N_{j}^{t-2}$. This shows that Eq.~\eqref{bigclaim} holds.
This is still not enough to conclude, as we need to show that over the full state $\mathbf{N}^{t-2}=\mathbf{N}^{t}$, or $\mathbf{N}^{t-1}=\mathbf{N}^{t}$. 
This is implied by the fact that if $N^t_j = N^{t-1}_j$ for some $j$ and $t$ large enough to have $\Delta E=0$
\begin{equation}
    N^{t+1}_j = N^{t}_j
    \quad\text{or}\quad
    N^{t+1}_j = N^{t-1}_j = N^t_j
\end{equation}
using Eq.~\eqref{bigclaim} and  $N^t_j = N^{t-1}_j$. This implies that if a site $j$ is constant at any time update in the cycle, it will be constant along the whole cycle. All other non-constant sites are then prevented to ever satisfy $N^t_j = N^{t-1}_j$, hence by Eq.~\eqref{bigclaim} must satisfy $N^{t+1}_j = N^{t-1}_j \neq N^t_j$, leading overall to a 2-cycle. If all sites have a state constant in time instead, we have a fixed point.
\newline

\textbf{All fixed points in the uniform $d$-regular case of discrete gLV are characterized by extinction.}  It can happen that $S^*$ is characterized by the fact that for each $i\in S^*$, all the neighbors $j\in\partial i$ satisfy $j\notin S^*$. If this is the case, then the system is in an \textit{independent set fixed point}, simply because each surviving site $i\in S^*$ has state $N_i=K$, while the neighbors have $N_j=0$. Indeed, if all the neighbors are extinct, then $N_i$ must saturate to $K$. Furthermore, for a uniform gLV model on a $d$-regular graph for each $\alpha\notin \{\alpha^{*}_l\}_{l=1}^{l_{\rm max}}$, the only fixed points that exist are these independent set fixed points. This follows trivially from the fact that at a fixed point $g_i=0$ $\forall i$. This means that either $N_i=0$ or all the $K-N_i-\alpha\sum_{j\in\partial i} N_j=0$ (the second option is not possible for $\alpha\notin \{\alpha^{*}_l\}_{l=1}^{l_{\rm max}}$ unless $N_j = 0$ for all $j \in \partial i$). So all fixed points in the uniform $d$-regular case are characterized by extinction.

\section{Details about the theoretical methodology and notes about numerical implementation of BDCM equations}\label{BP_theory_appendix}
We now present the technical methods  used in our analysis, in particular we consider the Replica Symmetric (RS) version of the Backtracking Dynamical Cavity Method, we detail the fully connected limit, and how it reconnects to percolation. This method was originally developed to study zero temperature dynamics of the Ising model and graph cellular automata, and the full details can be found in \cite{BDCM, CellAuto}.
\subsection{The Backtracking Dynamical Cavity Method}\label{BDCM}
\textbf{Detailed expressions of the probability distributions needed to compute $\alpha_{\rm atyp}$, $\alpha_{\rm ext}$ and observables.}
We now detail how to compute the entropy $\Phi_{(p/c)}$ and the average values of the observables with respect to the following two probability distributions, adopting the same procedure originally introduced in \cite{BDCM,CellAuto}. 

First, to emphasize the locality of the update rule (i.e., the state of each site depends only on the state of its neighbors and the site state itself), we define
\begin{equation}\label{Rdef}
\mathcal{F}_i(\mathbf{N}) = \mathcal{R}\bigl(N_i,\{N_j\}_{j\in\partial i}\bigr),
\end{equation}

where $\{N_j\}_{j\in\partial i}$ are the biomasses of the neighboring sites of $i$. We consider the following probability distributions:
\begin{equation}
\label{verylongequation}
\begin{aligned}
\mathbbm{P}(\{\underline{N}_i\}_{i=1}^S)
&= \frac{1}{\mathcal{Z}_{(p/c)}(G,\alpha_{ij})}
\prod_{i=1}^{S}\mathbbm{1}\!\left[N_i^{t=1}>0\right] \\
&\quad \times \prod_{i=1}^S
\mathbbm{1}\!\left[
N_i^{p+1}=\mathcal{R}\!\left(N_i^{p+c},\{N_j^{p+c}\}_{j\in\partial i}\right)
\right] \\
&\quad \times \prod_{i=1}^S \prod_{t=1}^{p+c-1}
\mathbbm{1}\!\left[
N_i^{t+1}=\mathcal{R}\!\left(N_i^t,\{N_j^t\}_{j\in\partial i}\right)
\right],\\
\mathbbm{P}^{\rm Full.Occupied}(\{\underline{N}_i\}_{i=1}^S)
&= \frac{1}{\mathcal{Z}^{\rm Full.Occupied}_{(p/c)}(G,\alpha_{ij})}
\prod_{i=1}^{S}\mathbbm{1}\!\left[N_i^{t=1}>0\right]\mathbbm{1}\!\left[N_i^{t=p+c}>0\right] \\
&\quad \times \prod_{i=1}^S
\mathbbm{1}\!\left[
N_i^{p+1}=\mathcal{R}\!\left(N_i^{p+c},\{N_j^{p+c}\}_{j\in\partial i}\right)
\right] \\
&\quad \times \prod_{i=1}^S \prod_{t=1}^{p+c-1}
\mathbbm{1}\!\left[
N_i^{t+1}=\mathcal{R}\!\left(N_i^t,\{N_j^t\}_{j\in\partial i}\right)
\right].
\end{aligned}
\end{equation}

The first distribution coincides with Eq.~\eqref{longequation} and describes $(p/c)$-backtracking attractors reached from typical initializations (i.e., it selects the dominant attractor). The second distribution enforces the absence of extinction and is used to probe whether atypical initializations can lead to a fully occupied community even for $\alpha \geq \alpha_{\rm ext}$ (thus defining $\alpha_{\rm atyp}$).
To compute instead the number of (typical) attractors, it is sufficient to consider the first probability distribution at $p=0$ and without the constraint $\prod_{i=1}^S \mathbbm{1}[N_i^{t=1}>0]$.

We now present the theory for a general probability distribution of the form
\begin{equation}\label{C2}
\mathbbm{P}(\{\underline{N}_i\}_{i=1}^S)
= \frac{1}{\mathcal{Z}_{(p/c)}} \prod_{i=1}^S \mathcal{A}_{i}(\underline{N}_{i},\{\underline{N}_j\}_{j\in\partial i}),
\end{equation}
where $\mathcal{A}_{i}(\underline{N}_{i},\{\underline{N}_j\}_{j\in\partial i})$ is the \textit{constraint matrix}, which encompasses all distributions in Eq.~\eqref{verylongequation}.

\textbf{Observables, entropies and Message Passage BDCM recursion for a general topology and general interaction strengths. }
In order to compute the entropy density and the average value of local observables, such as $\eta_l$ and $\rho_0$ (where an observable $\Xi(\mathbf{N})$ is local provided that $\Xi(\mathbf{N})=\frac{1}{S}\sum_{i=1}^S\Xi_i(\underline{N}_i,\{\underline{N}_j\}_{j\in\partial i})$), we tilt the measure in Eq.~\eqref{C2} by the exponential weight $\prod_{i}e^{\lambda \Xi_i(\underline{N}_i,\{\underline{N}_j\}_{j\in\partial i})}$. Then the average value of $\Xi$ is given by
\begin{equation}\label{C3}
    \langle \Xi\rangle=\frac{\partial \Phi_{(p/c)}(\lambda)}{\partial \lambda}\Biggr|_{\lambda=0},\quad  
    \Phi_{(p/c)}(\lambda)=\frac{1}{S}\log\Bigl[\sum_{\{\underline{N}_i\}_{i=1}^S}\prod_{i=1}^S \mathcal{A}_{i}(\underline{N}_{i},\{\underline{N}_j\}_{j\in\partial i})e^{\lambda \Xi_i(\underline{N}_i,\{\underline{N}_j\}_{j\in\partial i})}\Bigr].
\end{equation}

Note that $\Phi_{(p/c)}(\lambda=0)=\frac{1}{S}\log(\mathcal{Z}_{(p/c)})$ is the entropy (density) of interest. 

\begin{figure}
    \centering
    \includegraphics[width=0.7\textwidth]{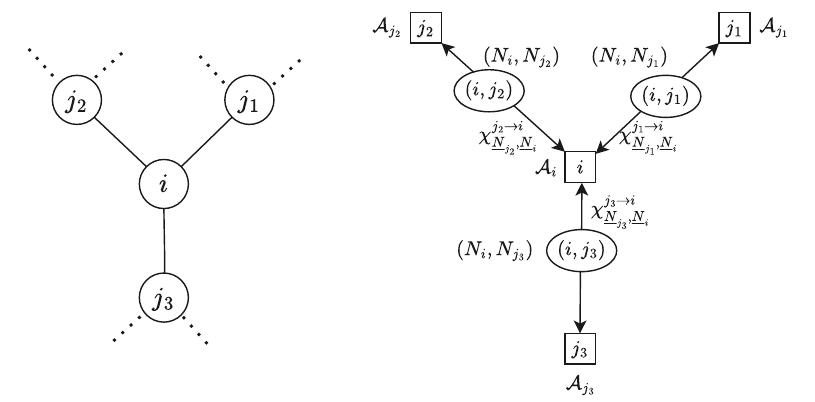}
    \caption{A graphical representation of the edge dual factor graph, with the factor nodes as squares and the variable nodes as ovals. $\mathcal{A}_i$ is a short-hand to denote $\mathcal{A}_{i}(\underline{N}_{i},\{\underline{N}_j\}_{j\in\partial i})
\, e^{\lambda \Xi_i(\underline{N}_i,\{\underline{N}_j\}_{j\in\partial i})}$. Note that the underlying topology of the edge dual and the original graph is the same. The variable nodes take values $(\underline{N}_i,\underline{N}_j)$. When  the product over the factors is taken the probability of interest is recovered.}
    \label{dual}
\end{figure}

We adopt the BDCM approach by representing the probability distribution using an edge-dual factor graph. In particular, for each pair of neighboring sites $i$ and $j$ in the original graph, we introduce a \textit{variable node} $(i,j)$ carrying the variables $(\underline{N}_i,\underline{N}_j)$. We also introduce a \textit{factor node} $(i)$ for each site $i$ of the original graph. The factor node encodes both the constraint and the tilt through a term of the form
\begin{equation}
\mathcal{A}_{i}(\underline{N}_{i},\{\underline{N}_j\}_{j\in\partial i})
\, e^{\lambda \Xi_i(\underline{N}_i,\{\underline{N}_j\}_{j\in\partial i})},
\end{equation}
which depends on the value of the neighboring variable nodes.
Figure~\ref{dual} shows the corresponding factor-graph representation of this probability distribution. The product over all the factor nodes reconstructs the probability measure of interest. This edge-dual factor graph is locally tree-like if the underlying topology is also locally tree-like, and therefore we apply Belief Propagation to compute the entropy and, consequently, the observables' averages (this is the Bethe assumption \cite{InfoPhysComp}).   

Indeed, for a factor graph of this type it has been shown in \cite{BDCM,CellAuto,CedricCounting,jankola2026minority} that the $\Phi$ (also called the \textit{free entropy}) can be computed as
\begin{align}\label{freeentropy}
    \Phi(\lambda)&=\frac{1}{S}\sum_{i=1}^S \log(Z_i)-\frac{1}{S}\sum_{i,j}\log(Z_{ij}),  \\
    Z_i&=\sum_{\underline{N}_i,\{\underline{N}_j\}_{j\in\partial i}}\mathcal{A}_{i}(\underline{N}_{i},\{\underline{N}_j\}_{j\in\partial i})
\, e^{\lambda \Xi_i(\underline{N}_i,\{\underline{N}_j\}_{j\in\partial i})}\prod _{k\in\partial i}\chi^{k\to i}_{\underline{N}_k,\underline{N}_i},\\
    Z_{ij}&=\sum_{\underline{N}_i,\underline{N}_j}\chi^{i\to j}_{\underline{N}_i,\underline{N}_j}\chi^{j\to i}_{\underline{N}_j,\underline{N}_i}.
\end{align}
The $\chi^{i\to j}_{\underline{N}_i,\underline{N}_j}$ are the (edge-dual) \textit{Belief Propagation messages}, which are given by the recursion relations
\begin{equation}\label{eqrec}
    \chi^{i\rightarrow j}_{\underline{N}_i,\underline{N}_j}=\frac{1}{Z^{\rightarrow}}\sum_{\{\underline{N}_h\}_{h\in\partial i/j}}{\mathcal{A}_{i}(\underline{N}_{i},\{\underline{N}_h\}_{h\in\partial i/j}\cup \underline{N}_j)
\, e^{\lambda \Xi_i(\underline{N}_i,\{\underline{N}_h\}_{h\in\partial i})}\prod_{h\in\partial i/j}\chi^{h\to i}_{\underline{N}_h,\underline{N}_i}}.
\end{equation}

We note that the recursion in Eq.~\eqref{eqrec} is precisely the recursion that the messages must satisfy so that the entropy in Eq.~\eqref{freeentropy} is extremized (i.e. the ``variational" derivative, which in this case is just the derivative w.r.t. the messages, of the free entropy is zero). Then, to compute the average values of observables, one can take the derivative with respect to $\lambda$ analytically (and evaluate it at $\lambda=0$). Indeed, one has that 
\begin{equation}\label{avarage}
    \langle \Xi\rangle = \frac{1}{S}\sum_i\frac{\sum_{\underline{N}_i,\{\underline{N}_j\}_{j\in\partial i}}\Xi_i(\underline{N}_i,\{\underline{N}_j\}_{j\in\partial i})\mathcal{A}_{i}(\underline{N}_{i},\{\underline{N}_j\}_{j\in\partial i})
\, \prod _{k\in\partial i}\chi^{k\to i}_{\underline{N}_k,\underline{N}_i}}{\sum_{\underline{N}_i,\{\underline{N}_j\}_{j\in\partial i}}\mathcal{A}_{i}(\underline{N}_{i},\{\underline{N}_j\}_{j\in\partial i})
\, \prod _{k\in\partial i}\chi^{k\to i}_{\underline{N}_k,\underline{N}_i}}.
\end{equation}

From Belief Propagation, one can also compute the marginals over the variable nodes, i.e., the probability distribution that $(\underline{N}_i,\underline{N}_j)$ takes a certain value. These are given by:
\begin{equation}
    \mu(\underline{N}_i,\underline{N}_j)=\frac{1}{Z_{ij}}\chi^{i\to j}_{\underline{N}_i,\underline{N}_j}\chi^{j\to i}_{\underline{N}_j,\underline{N}_i}.
\end{equation}

From the marginal probability distribution over factor nodes we derive the probability that a site $i$ is occupied, and the conditional probability that site $i$ is occupied given that site $j$ is. 
\begin{equation}\label{C10}
    \mu_i=\sum_{\underline{N}_i,\underline{N}_j}\mu(\underline{N}_i,\underline{N}_j)\indi\Bigl[N_{i}^{p+c}>0\Bigr], \quad 
    \mu_{i|j}=\frac{\sum_{\underline{N}_i,\underline{N}_j}\mu(\underline{N}_i,\underline{N}_j)\indi\Bigl[N_{i}^{p+c}>0\Bigr]\indi\Bigl[N_{j}^{p+c}>0\Bigr]}{\mu_j}.
\end{equation}
An additional object that will be important in the percolation analysis will be the probability $P\left(\{\underline N_u\}_{u\in\partial i\setminus j} \,\middle|\, \underline N_{i},\underline N_j\right) $, denoting the probability that the neighbors $u\in\partial i\setminus j$ of site $i$ follow the trajectories $\underline{N}_u$, assuming that site $i$ and $j\in\partial i$ ($j$ is a neighbor of $i$ that has been singled out) follow the trajectories $\underline{N}_i$ and $\underline{N}_j$. In particular it holds that 
\begin{align}\label{E19A} P\left(\{\underline N_u\}_{u\in\partial i/ j} \,\middle|\, \underline N_{i},\underline N_j\right) &= \frac{\mathcal{A}_{i}\left(\underline N_{i},\underline N_j\cup \{\underline N_u\}_{u\in\partial i/ j}\right) \prod_{u\in\partial i/ j} \chi^{u\to i}_{\underline N_u,\underline N_{i}}}{\sum_{\{\underline N'_u\}_{u\in\partial i/ j}} \mathcal{A}\left(\underline N_{i},\underline N_j\cup \{\underline N'_u\}_{u\in\partial i/ j}\right) \prod_{u\in\partial i/ j} \chi^{u\to i}_{\underline N'_u,\underline N_{i}}}. \end{align}
\textbf{Specialized BDCM equations for a $d$-regular graph.}
While solving these equations for a general graph involves sums over a total of $O(SK^{d(p+c)})$ states, as opposed to the sum over $O(K^S)$ needed to compute the expression in Eq.~\eqref{C3}, this complexity can be further reduced in the case of $d$-regular graphs with uniform interaction. Indeed, all the messages can be assumed homogeneous (in the Replica Symmetric assumption), and therefore we can drop the explicit dependency $i\to j$. Thus, the fixed-point update equation for the messages becomes
\begin{equation}\label{C11}
   \chi_{\underline{N},\underline{M}}^{\rightarrow}=\frac{1}{Z^{\rightarrow}}\sum_{\{\underline{y}_i\}_{i=1}^{d-1}}{\mathcal{A}(\underline{N}, \{\underline{y}_i\}_{i=1}^{d-1} \cup \underline{M})\prod_{i=1}^{d-1}\chi_{\underline{y}_i,\underline{N}}^{\rightarrow}},
\end{equation}
and the entropy and the expression for the average values of observables are also simplified considerably:
\begin{align}
    \Phi(\lambda)&=\log(Z_{\rm fac})-\frac{d}{2}\log(Z_{\rm var}),\\
    Z_{\rm fac}&=\sum_{\underline{N},\{\underline{y}\}_{i=1}^d}e^{\lambda \Xi(\underline{N},\{\underline{y}\}_{i=1}^d)}\mathcal{A}(\underline{N},\{\underline{y}\}_{i=1}^d)\prod_{i=1}^d\chi_{\underline{y}_i,\underline{N}}^{\rightarrow},\\
    Z_{\rm var} & = \sum_{\underline{N},\underline{M}}\chi_{\underline{N},\underline{M}}^{\to}\chi_{\underline{M},\underline{N}}^{\to},\\
    \label{C15}
    \langle \Xi\rangle &= \frac{\sum_{\underline{N},\{\underline{y}_i\}_{i=1}^d}\Xi(\underline{N},\{\underline{y}_i\}_{i=1}^d)\mathcal{A}(\underline{N},\{\underline{y}_i\}_{i=1}^d)
\, \prod _{i=1}^d\chi^{\to}_{\underline{y}_i,\underline{N}}}{\sum_{\underline{N},\{\underline{y}_i\}_{i=1}^d}\mathcal{A}(\underline{N},\{\underline{y}_i\}_{i=1}^d)
\, \prod _{i=1}^d\chi^{\to}_{\underline{y}_i,\underline{N}}}.
\end{align}
Solving the recursion in Eq.~\eqref{C11} still involves a sum over $O(K^{d(p+c)})$ states, we detail in Appendix~\ref{FFT} how to remove the exponential dependency on $d$, reducing the equations to sums over $O(dK^{(p+c)})$ states, which can be computed efficiently. This set of equations coincides with Eq.~\eqref{main:begin} in the main text.
\subsection{High degree limit of the BDCM equations}\label{high_d_limit}
\textbf{Derivation of the expression of the messages in the high degree limit.}
We derive now the high degree limit of the BDCM equations. For generality, let's consider $\alpha_{ij}$ to be i.i.d. with variance $\sigma/\sqrt d$ and expectation $\alpha/d$, for a $d$-regular graph. The goal is to take the limit $d\to\infty$, and we will later focus on the case $\sigma=0$ (uniform interactions). For a fixed realization of the $\alpha_{ij}$ the messages (at $\lambda=0$) read
\begin{equation}\label{D17}
    \chi^{i\rightarrow j}_{\underline{N}_i,\underline{N}_j}=\frac{1}{Z^{\rightarrow}}\sum_{\{\underline{N}_h\}_{h\in\partial i/j}}{\mathcal{A}(\underline{N_i}, \{\underline{N}_h\}_{h\in\partial i};\{\alpha_{ih}\}_{h\in\partial i})\prod_{h\in\partial i/j}\chi_{\underline{N_h},\underline{N_i}}^{h\rightarrow i}}, 
\end{equation}
and $\mathcal{A}(\underline{N_i}, \{\underline{N}_h\}_{h\in\partial i};\{\alpha_{ih}\}_{h\in\partial i})=g(\underline{N}_i, \underline{N}_j \alpha_{ij}+\sum_{h\in\partial i/j}\underline{N}_h\alpha_{ih})$, for some function $g$ (given the type of dynamic considered in Eq.\eqref{discrete_glv}). Then, introducing the cavity field $\underline{S}^{i\to j}=\sum_{h\in\partial i/j}\underline{N}_h\alpha_{ih}$, we can write
\begin{equation}
    \chi^{i\rightarrow j}_{\underline{N}_i,\underline{N}_j}=\frac{1}{Z^{\rightarrow}}\int d\underline{S}^{i\to j}g(\underline{N}_i,\underline{N}_j\alpha_{ij}+\underline{S}^{i\to j})\sum_{\{\underline{N}_h\}_{h\in\partial i/j}}\prod_{h\in\partial i/j}\chi_{\underline{N_h},\underline{N_i}}^{\rightarrow}\delta(\underline{S}^{i\to j}-\sum_{h\in\partial i/j}\underline{N}_h\alpha_{ih}). 
\end{equation}
The term $\sum_{\{\underline{N}_h\}_{h\in\partial i/j}}\prod_{h\in\partial i/j}\chi_{\underline{N_h},\underline{N_i}}^{\rightarrow}\delta(\underline{S}^{i\to j}-\sum_{h\in\partial i/j}\underline{N}_h\alpha_{ih})$ is just the probability distribution of the cavity fields. 
In the $d \gg 1$ limit, the factor graph is no longer loop-free. Nevertheless, correlations between cavity messages can be discarded \cite{InfoPhysComp,Review1} (this is the well-known r-BP limit). Due to the Central Limit Theorem the probability distribution of $\underline{S}^{i\to j}$  will be a Gaussian, and thus:
\begin{equation}
    \chi^{i\rightarrow j}_{\underline{N}_i,\underline{N}_j}=\frac{1}{Z^{\rightarrow}}\int d\underline{S}^{i\to j}g(\underline{N}_i,\underline{N}_j\alpha_{ij}+\underline{S}^{i\to j})\mathcal{N}^{i\to j}(\underline{S}^{i\to j}|\underline{m}^{i\to j},\Sigma_{i\to j}),
\end{equation}
where $\Sigma_{i\to j}\in \mathbb{R}^{T\times T}$ is the covariance matrix ($T=p+c$, the total time horizon). We notice that $\alpha_{ij}\propto 1/d$, meaning that $\underline{N}_j$ does not influence the messages. Then,  $\chi^{i\rightarrow j}_{\underline{N}_i,\underline{N}_j}=\psi_i(\underline{N}_i)/Z$ for some (normalized) $\psi_i$ for every $\underline{N}_j$ (thus $Z=(K+1)^T$). This $\psi_i$ is the marginal over $\underline{N}_i$. Indeed, the edge dual marginal is $\mu_{ij}(\underline{N}_i,\underline{N}_j)\propto \chi^{i\rightarrow j}_{\underline{N}_i,\underline{N}_j}\chi^{j\rightarrow i}_{\underline{N}_j,\underline{N}_i}\propto \psi_i(\underline{N}_i)\psi_j(\underline{N}_j)$. Additionally, the covariance matrix is given by
\begin{equation}
\begin{aligned}
\Sigma_{tt'}^{i\to j}
&= \mathbb{E}_{\{\underline{N}_h \sim \psi_h\}_{h \in \partial i \setminus j}}
\Biggl[
\Bigl(\sum_{h \in \partial i \setminus j} \alpha_{ih} N_h^t \Bigr)
\Bigl(\sum_{h \in \partial i \setminus j} \alpha_{ih} N_h^{t'} \Bigr)
\Biggr] \\
&\quad
- \mathbb{E}_{\{\underline{N}_h \sim \psi_h\}_{h \in \partial i \setminus j}}
\Biggl[
\sum_{h \in \partial i \setminus j} \alpha_{ih} N_h^t
\Biggr]
\mathbb{E}_{\{\underline{N}_h \sim \psi_h\}_{h \in \partial i \setminus j}}
\Biggl[
\sum_{h \in \partial i \setminus j} \alpha_{ih} N_h^{t'}
\Biggr] \\
&= \sum_{h \in \partial i \setminus j} \alpha_{ih}^2
\Bigl(
\mathbb{E}_{\underline{N}_h \sim \psi_h}\bigl[N_h^t N_h^{t'}\bigr]
- \mathbb{E}_{\underline{N}_h \sim \psi_h}\bigl[N_h^t\bigr]
  \mathbb{E}_{\underline{N}_h \sim \psi_h}\bigl[N_h^{t'}\bigr]
\Bigr),
\end{aligned}
\end{equation}
and for $\sigma=0$ this converges to zero in the limit $d\gg1$. Then the mean vector is just:
\begin{equation}
    \underline{m}^{i\to j}=\sum_{h\in\partial i/j}\alpha_{ih}\mathbb{E}_{\underline{N}_h \sim \psi_h}\Bigl[\underline{N}_h\Bigr]
\end{equation}
In the case of uniform $\sigma=0$ interactions all the messages are thus equivalent, and there is no explicit dependence on $i$ (or $j$). Thus $\underline{m}^{i\to j}=\alpha\mathbb{E}_{\underline{M} \sim \psi}\bigl[\underline{M}]$, and $\Sigma_{tt'}^{i\to j}=0$. The $\psi(\underline{N})$ (note that we lost the dependence on $i$) satisfy the following recursion
\begin{equation}\label{C23}
    \psi(\underline{N})\propto g\bigr (\underline{N},\alpha\mathbb{E}_{\underline{M} \sim \psi}\bigl[\underline{M}\bigr] \bigl),
\end{equation}
This is the final step of the $d\to \infty$ limit. We went from Eq.~\eqref{D17} for the messages $\chi_{\underline{N}_i\underline{N}_j}$, which still depended  on $d$, to a  reduced equation for the quantity $\psi(\underline{N})$, which is the marginal in the fully connected limit. 

\textbf{Extracting observables and thresholds in the limit $d\gg1 $. }
We now detail how we computed the $\alpha_{\rm ext}$ and $\alpha_{\rm atyp}$ in the limit $d\to \infty$. We start from the easier computation of $\alpha_{\rm ext}$. To compute these quantities the first step is to solve Eq.~\eqref{C23}. This is done numerically with a fixed point solver, we detail specifically how to do this for $\alpha_{\rm ext}$ and $\alpha_{\rm atyp}$ separately.

\textit{a) Computing the fraction $\rho_0$ and the $\alpha_{\rm ext}$ threshold.}
In the case of the fully connected limit solving Eq.~\eqref{C23} is much simpler compared to equations Eq.~\eqref{C11}, as there is no dependence on $d$, we need to track only the trajectory of a single site, and there is no need to compute sum and products between the various messages. This means that we are able to consider $T$ up to ten even at $K=4$. Furthermore, we find (more on this below) that in the fully connected limit the convergence to the steady state is very fast (it turns out that the dynamic converges in 2 or 3 steps even at $K=4$).  Thus, to determine $\alpha_{\rm ext}$ the full BDCM formalism is unnecessary, and we can consider the $g$ function without any backtracking term. In other words, we can consider just the forward dynamic, without conditioning on the final attractor, as this will be reached in very few steps. To do this, $g$ has to be the following:
\begin{equation}
g(\underline{N},\underline{M})
=\mathbbm{1}[N^{t=1}>0]\prod_{t=1}^{T}
\mathbbm{1}\!\left[N^{t+1}=\mathcal{R}(N^t,M^t)\right],
\end{equation}
where $\mathcal{R}$ has been defined in Eq.~\eqref{Rdef} (losing the explicit dependency on $i,j$). The average value of $\rho_0$ at time $t$ is given by 
\begin{equation}\label{C24}
\rho_0^t=\sum_{\underline{M}}\psi(\underline{M})\,\mathbbm{1}[M_t=0],
\end{equation}
given that $\psi$ are just the marginals. Thus, we solve Eq.~\eqref{C23} at $T=10$, track $\rho_0^t$, note that it converges in few steps (as said before, for $K=4$ in much less than 10 steps), and take the converged value as the fraction of vacant sites. This is what is plotted in Figure~\ref{high_degree_K_2} and Figure~\ref{high_degree_K_3} (with a numerical comparison).

\textit{ b) Extracting the $\alpha_{\rm atyp}$ threshold and the $\Phi^{\rm Full.Occupied}_{(p/c)}$ entropy.}
The entropy in the fully connected limit is given by
\begin{equation}\label{C25}
    \Phi=\log\Bigl(\sum_{\underline{N}} g\bigr (\underline{N},\alpha\mathbb{E}_{\underline{M} \sim \psi}\bigl[\underline{M}\bigr] \bigl)\Bigr).
\end{equation}

To compute the $\alpha_{\rm atyp}$ threshold in this $d\to \infty$ limit, we need to use the full BDCM formalism again,  including terms that prevent extinction. Thus, the $g$ function must be written as:
\begin{equation}\label{C27}
    g(\underline{N},\underline{M})^{}
=\mathbbm{1}[N^{t=1}>0]\mathbbm{1}[N^{t=p+c}>0]\mathbbm{1}[N^{t=p+1}=\mathcal{R}(N^{p+c},M^{p+c})]\prod_{t=1}^{p+c-1}
\mathbbm{1}\!\left[N^{t+1}=\mathcal{R}(N^t,M^t)\right] .
\end{equation}
Then, we compute the associated entropy $\Phi^{\rm Full.Occupied}_{(p/c)}$, and the $\alpha$ value for which it becomes negative is the critical threshold $\alpha_{\rm atyp}^{d\to \infty}$. This threshold remains stable for all $p$ values considered.
See Figure~\ref{figure_d_infty_atyp} for $\Phi^{\rm Full.Occupied}_{(p/c)}$ and $\alpha_{\rm atyp}^{d\to \infty}$.

\subsection{Message passing for dynamics-dependent percolation}\label{Percolation_Exact}
In this section, we develop the general message passing approach that we use to study the percolation of an attractor of the gLV dynamic. We stress from the start that the equations derived below can be applied to any dynamical system, including stochastic ones, whose underlying probability measure admits a locally tree-like factor-graph representation of the type shown in Figure~\ref{dual}
(see \cite{InfoPhysComp} for a general definition of factor graphs).

\textbf{Preliminary definitions and notation.} We will firstly develop our theory assuming that the original graph $G=(V,E)$ is a tree. Once we find equations for that case we will be able to argue, by virtue of the Bethe approximation, that the formula we find is correct in the thermodynamic limit ($S\gg1$) even on locally tree-like random graphs (assuming the replica symmetric assumption for the gLV model is correct). This is the standard approach when developing message-passing algorithms, see \cite{InfoPhysComp} for the full details. 

Let site $i\in V$ have $d(i)$ neighbors, denoted by $i_1,\dots,i_{d(i)}$. For each $\ell=1,\dots,d(i)$, consider the \textit{cavity graphs} rooted in $i_\ell$, obtained by removing the edge $(i,i_\ell)$.  These cavity graphs will be all disconnected from each other, provided that $G$ is a tree (we remind that a tree is a graph in which any two sites are connected by a unique path). The precise definition for the set of sites belonging to the cavity graphs, which we denote  by $T^{i_\ell}_{(i,i_\ell)}$, is the following:
\begin{equation}
    T^{i_\ell}_{(i,i_\ell)}=\Bigl\{k\in V\setminus \{i\} \,\,\text{ such that } \,\, \exists v_0,...,v_m\in V\setminus \{i\} \text{ with } v_0=i_{\ell},\,\, v_m=k, \, (v_{r-1},v_r)\in E,\,\, \forall r=1,\dots,m \Bigr\},
\end{equation}
and if $G$ is a tree $T^{i_\ell}_{(i,i_\ell)}\cap T^{i_{\ell'}}_{(i,i_{\ell'})} =\emptyset$, for each $\ell\neq \ell'$. The edges of the cavity graphs are the edges in $G$ restricted to  the sites in $ T^{i_\ell}_{(i,i_\ell)}$.

Let $a(\underline N_i)$ be the indicator function that returns one if the trajectory $\underline N_i$ corresponds to a site that is occupied at convergence of the dynamics, and zero otherwise. We write
\begin{equation}
    A_i \equiv a(\underline N_i)=\indi\!\left[N_i^{t=p+c}>0\right].
\end{equation}

Given a site $i$, we define its occupied cluster $\mathcal C_i$ as
\begin{equation}
    \mathcal C_i=\Bigl\{k\in V:\exists\, m\geq 0,\ \exists\, v_0,\dots,v_m\in V\ \text{such that}\ v_0=i,\ v_m=k,\ (v_{r-1},v_r)\in E,\ A_{v_r}=1\ \forall r=1,\dots,m\Bigr\}.
\end{equation}
In words, $\mathcal C_i$ is the connected component of occupied sites containing $i$. If $i$ is not occupied, then $\mathcal C_i=\emptyset$. We denote its size by
\begin{equation}
    S_i=|\mathcal C_i|.
\end{equation}
We also define the cavity clusters $\mathcal C^{i_\ell\to i}$ relative to the cavity graphs:
\begin{equation}
    \mathcal C^{i_\ell\to i}=\Bigl\{k\in T^{i_\ell}_{(i,i_\ell)}:\exists\, m\geq 0,\ \exists\, v_0,\dots,v_m\in T^{i_\ell}_{(i,i_\ell)}\ \text{such that}\ v_0=i_\ell,\ v_m=k,\ (v_{r-1},v_r)\in E,\ A_{v_r}=1\ \forall r=1,\dots,m\Bigr\}.
\end{equation}
We denote their cardinalities by
\begin{equation}
    S^{i_\ell\to i}=|\mathcal C^{i_\ell\to i}|.
\end{equation}
With these definitions, the size of the occupied cluster containing $i$ satisfies
\begin{equation}\label{E6}
    S_i=\begin{cases}0, & \text{if } A_i=0,\\ 1+\sum_{\ell=1}^{d(i)}S^{i_\ell\to i}, & \text{if } A_i=1.\end{cases}
\end{equation}
Similarly, for a cavity cluster $\mathcal C^{i_\ell\to i}$ (rooted at $i_\ell$), one has
\begin{equation}\label{E7}
    S^{i_\ell\to i}=\begin{cases}0, & \text{if } A_{i_\ell}=0,\\ 1+\sum_{u\in\partial i_\ell\setminus i}S^{u\to i_\ell}, & \text{if } A_{i_\ell}=1,\end{cases}
\end{equation}
where $\partial i_\ell\setminus i$ denotes the set of neighbors of $i_\ell$ excluding $i$, and $S^{u\to i_\ell}$ is the size of the corresponding cavity cluster $\mathcal{C}^{u\to i_{\ell}}$.

\textbf{Generating functions and fixed point recursions.} Since we are interested in cluster sizes, we introduce the following \textit{generating function} (the idea of considering generating functions to study network problems was brought to widespread use by Newman et al. in \cite{original_newman}):
\begin{equation}
    H^{i_\ell\to i}_{\underline N_{i_\ell},\underline N_i}(z)=\sum_{s=0}^{f(S)}\pi^{i_{\ell}\to i}\left(S^{i_\ell\to i}=s\,\middle|\,\underline N_{i_\ell},\underline N_i\right)z^s,
\end{equation}
where $\pi^{i_{\ell}\to i}$ is  the cavity probability that site $i_{\ell}$ belongs to a cluster of $s$ sites relatively to the cavity graphs $T^{i_{\ell}}_{i,i_{\ell}}$ once we condition on $\underline{N}_{i},\underline{N}_{i_\ell}$. $f(S)$ is the cutoff function, that for a finite $S$ tells up to which size a cluster is considered small. The precise value of $f(S)$ is not important, as once we take the limit $S\to \infty$ also $f(S)$ will go to infinity. 
In short, the $f(S)$ is needed to denote the fact that we are summing over small cluster sizes (this is also what is done implicitly in ~\cite{PercolationBP}). From now on we denote $\sum_{s=0}^{f(S)}=\overset{\odot}{\sum}_s$. Note that in general
\begin{equation}
    H^{i_\ell\to i}_{\underline N_{i_\ell},\underline N_i}(1)\neq 1,
\end{equation}
as the missing probability mass corresponds to the probability that the cavity cluster $\mathcal C^{i_\ell\to i}$ is larger than the cutoff. Therefore $H^{i_\ell\to i}_{\underline N_{i_\ell},\underline N_i}(1)$ is the conditional probability that the cavity cluster $\mathcal{C}^{i_{\ell}\to i}$ is small, given the boundary trajectories $(\underline N_{i_\ell},\underline N_i)$. We now derive a recursion for $H^{i_\ell\to i}_{\underline N_{i_\ell},\underline N_i}(z)$. We are going to repeatedly use the Bayes theorem, which states that for two events $A,B$ it holds that  \begin{equation}
    P(A,B)=P(A|B)P(B).
\end{equation}
The following holds by applying the Bayes theorem
\begin{equation}\label{E12}
     H^{i_\ell\to i}_{\underline N_{i_\ell},\underline N_i}(z)=\Ssum_s\sum_{\{\underline{N}_u\}_{u\in\partial i_{\ell}/ i}}\pi^{i_{\ell}\to i}\left(S^{i_\ell\to i}=s\,\middle|\,\underline N_{i_\ell},\underline N_i, \{\underline{N}_u\}_{u\in\partial i_{\ell}/ i} \right)z^sP(\{\underline{N}_u\}_{u\in\partial i_{\ell}/ i}|\underline N_{i_\ell},\underline N_i).
\end{equation}
where $\pi^{i_{\ell}\to i}\left(S^{i_\ell\to i}=s\,\middle|\,\underline N_{i_\ell},\underline N_i, \{\underline{N}_u\}_{u\in\partial i_{\ell}/ i} \right)$ is again a conditional probability for the sizes of cluster relative to the graph $T^{i_{\ell}}_{(i,i_{\ell})}$. The probability $\pi^{i_{\ell}\to i}$ can be written as
\begin{align}
&\pi^{i_{\ell}\to i}\left(S^{i_\ell\to i}=s\,\middle|\,\underline N_{i_\ell},\underline N_i, \{\underline{N}_u\}_{u\in\partial i_{\ell}/ i} \right) \notag \\
&\qquad =\pi^{i_{\ell}\to i}\left(S^{i_\ell\to i}=s,\,A_{i_\ell}=1\,\middle|\,\underline N_{i_\ell},\underline N_i,\{\underline{N}_u\}_{u\in\partial i_{\ell}/ i}\right)+\pi^{i_{\ell}\to i}\left(S^{i_\ell\to i}=s,\,A_{i_\ell}=0\,\middle|\,\underline N_{i_\ell},\underline N_i,\{\underline{N}_u\}_{u\in\partial i_{\ell}/ i}\right)\\
&\qquad =\pi^{i_{\ell}\to i}\left(S^{i_\ell\to i}=s\,\middle|A_{i_\ell}=1,\,\underline N_{i_\ell},\underline N_i,\{\underline{N}_u\}_{u\in\partial i_{\ell}/ i}\right)\delta_{1,a(\underline{N_{i_{\ell}}})} \notag \\
&\hspace{6cm} +\pi^{i_{\ell}\to i}\left(S^{i_\ell\to i}=s\,\middle|\,A_{i_\ell}=0,\,\underline N_{i_\ell},\underline N_i,\{\underline{N}_u\}_{u\in\partial i_{\ell}/ i}\right)\delta_{0,a(\underline{N_{i_{\ell}}}).}\label{D41}
\end{align}
We can now substitute the final expression above  in Eq.~\eqref{E12}, and apply the property in Eq.~\eqref{E7} to get
\begin{align}\label{app:D42} H^{i_\ell\to i}_{\underline N_{i_\ell},\underline N_i}(z) &= \delta_{0,a(\underline N_{i_\ell})} + z\,\delta_{1,a(\underline N_{i_\ell})}\sum_{\{\underline N_u\}_{u\in\partial i_{\ell}/ i}} \Ssum_{\{s_k\}_{k=1}^{d(i_{\ell})-1}}\Bigl(\prod_{k=1}^{d(i_\ell)-1} z^{s_k}\Bigr) \notag \\ &\qquad \times \pi^{\partial i_{\ell}\setminus i\to i_{\ell}}\left(\{S^{u\to i_{\ell}}=s_k\}_{\substack{
u\in\partial i_{\ell}\setminus i\\
k\in\{1,\dots,d(i_{\ell})-1\}
}} \,\middle|\, \underline N_{i_\ell},\underline N_i,\{\underline N_u\}_{u\in\partial i_{\ell}/ i}\right) P\left(\{\underline N_u\}_{u\in\partial i_{\ell}/ i} \,\middle|\, \underline N_{i_\ell},\underline N_i\right), \end{align}
where we denoted with $ \pi^{\partial i_{\ell}\setminus i\to i_{\ell}}$ the conditional joint probability for the sizes of the clusters $\{\mathcal C^{u\to i_{\ell}}\}_{u\in\partial i_{\ell}\setminus i}$.

\textbf{Dynamical conditioning on full edge state.}
The reason why we insert this conditioning argument in this work is that, with this conditioning, the conditional probability $\pi^{\partial i_{\ell}\setminus i\to i_{\ell}}(\cdot | \underline{N}_{i_{\ell}},\underline{N_i},\{\underline{N}_u\}_{u\in\partial i_{\ell}\setminus i})$ can be exactly factorized:
\begin{equation}\label{E8}
   \pi^{\partial i_{\ell}\setminus i\to i_{\ell}}\left(\{S^{u\to i_{\ell}}\}_{
u\in\partial i_{\ell}\setminus i} \,\middle|\, \underline N_{i_\ell},\underline N_i,\{\underline N_u\}_{u\in\partial i_{\ell}/ i}\right)=\prod_{u\in\partial i_{\ell}\setminus i}\pi^{u\to i_{\ell}}\left(S^{u\to i_\ell}\,\middle|\,\underline N_{i_\ell}, \underline{N}_i,\underline N_u\right).
\end{equation}
On a tree, this is an exact conditional-independence statement (on locally tree-like random graphs, it is asymptotically exact in the Bethe/cavity sense). This is true because in the edge-dual representation of the probability distribution in Eq.~\eqref{longequation} (see Figure~\ref{dual}), fixing the trajectories $(\underline N_i,\underline N_{i_\ell})$ separates the different branches of the edge-dual factor graph (and if $G$ is a tree also the edge dual factor graph is).
This is the key observation that allows us to formulate the percolation problem in terms of trajectory-resolved cavity messages. We remark that in previous work (see for instance \cite{Site_percolation,PercolationBP}) this conditioning was not needed, as those works studied cases in which the occupation of a site was independent of the others. In those cases, it is the joint unconditioned probability $\pi^{\partial i_{\ell}\setminus i\to i_{\ell}}(\{S^{u\to i_{\ell}}\})$ itself that factorizes, allowing to obtain much simpler recursions, something which does not happen in our case due to dynamically-induced correlations.

Thus, Eq.~\eqref{E8} allows us to write Eq.~\eqref{app:D42} as
\begin{align} H^{i_\ell\to i}_{\underline N_{i_\ell},\underline N_i}(z) &= \delta_{0,a(\underline N_{i_\ell})} + z\,\delta_{1,a(\underline N_{i_\ell})}\sum_{\{\underline N_u\}_{u\in\partial i_{\ell}/ i}} \prod_{u\in \partial i_{\ell}\setminus i} \Bigl(\Ssum_{s_k} z^{s_k} \notag \\ &\qquad \times \pi^{u\to i_{\ell}}\left(S^{u\to i_{\ell}}=s_k\,\middle|\, \underline N_{i_\ell},\underline N_i,\underline N_u\right)\Bigr) P\left(\{\underline N_u\}_{u\in\partial i_{\ell}/ i} \,\middle|\, \underline N_{i_\ell},\underline N_i\right). \end{align}
We note that $\pi^{u\to i_{\ell}}\left(S^{u\to i_{\ell}}=s_k\,\middle|\, \underline N_{i_\ell},\underline N_i,\underline N_u\right)=\pi^{u\to i_{\ell}}\left(S^{u\to i_{\ell}}=s_k\,\middle|\, \underline N_{i_\ell},\underline N_u\right)$, as once $\underline N_{i_\ell}$ and $\underline N_u$ have been specified the evolution of sites $i$ does not matter (again, due to the tree like structure of the factor graph). Then, we note that the object in parentheses is precisely the definition of $H^{u\to i_{\ell}}(\underline{N}_u,\underline{N}_{i_{\ell}})$. This allows us to write the following self-consistent recursion for the generating function:
\begin{align}\label{app:D45}
H^{i_\ell\to i}_{\underline N_{i_\ell},\underline N_i}(z) &= \delta_{0,a(\underline N_{i_\ell})} + z\,\delta_{1,a(\underline N_{i_\ell})}\sum_{\{\underline N_u\}_{u\in\partial i_{\ell}/ i}} \prod_{u\in \partial i_{\ell}\setminus i}H^{u\to i_{\ell}}_{\underline{N}_{u},\underline{N}_{i_{\ell}}}(z)P\left(\{\underline N_u\}_{u\in\partial i_{\ell}/ i} \,\middle|\, \underline N_{i_\ell},\underline N_i\right). \end{align}
The expression of $P\left(\{\underline N_u\}_{u\in\partial i_{\ell}/ i} \,\middle|\, \underline N_{i_\ell},\underline N_i\right)$ is given by Eq.~\eqref{E19A}, and follows directly from the BDCM method. So, the full specialized expression, depending  on the cavity messages $\chi$, is given by
\begin{align}\label{E19} 
H^{i_\ell\to i}_{\underline N_{i_\ell},\underline N_i}(z) &= \delta_{0,a(\underline N_{i_\ell})} + z\,\delta_{1,a(\underline N_{i_\ell})}\sum_{\{\underline N_u\}_{u\in\partial i_{\ell}/ i}}\frac{\mathcal{A}_{i_\ell}\left(\underline N_{i_\ell},\underline N_i\cup\{\underline N_u\}_{u\in\partial i_{\ell}/ i}\right) \prod_{u\in\partial i_{\ell}/ i} H^{u\to i_\ell}_{\underline{N}_u,\underline{N}_{i_\ell}}(z)\chi^{u\to i_\ell}_{\underline N_u,\underline N_{i_\ell}}}{\sum_{\{\underline N'_u\}_{u\in\partial i_{\ell}/ i}} \mathcal{A}_{i_{\ell}}\left(\underline N_{i_\ell},\underline N_i\cup \{\underline N'_u\}_{u\in\partial i_{\ell}/ i}\right) \prod_{u\in\partial i_{\ell}/ i} \chi^{u\to i_\ell}_{\underline N'_u,\underline N_{i_\ell}}}.\end{align}
As stated already, this equation is exact on a tree. We argue that this expression is also asymptotically exact in the limit $S\gg 1$ in the case of a tree-like factor graph, i.e., graphs for which there are very long loops whose length diverges with the system size, and assuming replica symmetry. The same assumption is needed to derive Eq.~\eqref{E19A} from BDCM, and it is equivalent to considering independent conditional probabilities $\pi^{u\to i_{\ell}}(s|\underline{N}_i,\underline{N}_{i_{\ell}})$. 
This is thus the usual Bethe approximation, and is what allows us to argue that the expression Eq.~\eqref{app:D45} is asymptotically exact for all locally tree-like topologies $G$ (and locally tree-like factor graphs). Once we specialize Eq.~\eqref{app:D45}  to the uniform $d$-regular case,  by noting that in that case the generating functions can be taken uniform (i.e.  $H^{i_\ell\to i}_{\underline{N},\underline{M}}(z)=H_{\underline{N},\underline{M}}$ for each $i,\ell$), we obtain Eq.~\eqref{recursion_MAIN} reported in the main. 

\textbf{Computing $\phi_{LC}$, the fraction of sites in the largest connected component.} We now need to relate the generating functions to $\phi_{\rm LC}$, the fraction of all sites belonging to the largest connected component of occupied sites. Here, we will rely on a common property of percolation, where the percolating cluster is unique, and its size can hence be expressed in terms of the complement of what is covered by the small components. 
In particular, to determine the size of the percolating cluster, we only need the value of $H^{i_\ell\to i}_{\underline N_{i_\ell},\underline N_i}(1)$, as this is the conditional probability that $\mathcal{C}^{i_{\ell}\to i}$ is small.
The reason why we wrote the equation for a general $z$ is that the full expression of $H^{i_\ell\to i}_{\underline N_{i_\ell},\underline N_i}(z)$ can be used to compute other interesting observables, such as the typical size of small components, by taking derivatives of the generating function with respect to $z$. See~\cite{PercolationBP} for more details.  

 We have that $\phi_{LC}$ is given by
\begin{equation}
    \phi_{LC}
    =
    \frac{1}{S}\sum_i
    \Bigl(
    \mathbb P[\text{$i$ is occupied}]
    -
    \mathbb P[\text{$i$ belongs to a small cluster $\mathcal C_i$ and $i$ is occupied}]
    \Bigr).
\end{equation}
If this fraction vanishes as $S\to\infty$, the system is in a non-percolating phase, since all occupied clusters are small. If instead $\phi_{\rm LC}>0$, the system is in a percolating phase. Equivalently, $\phi_{\rm LC}$ is the probability that a uniformly chosen site is occupied and belongs to an extensive cluster.

We will now express the corresponding probabilities via the message passing equations from the previous section, implicitly assuming their exactness in the thermodynamic limit $S\to \infty$. The probability that $i$ is occupied is simply the probability $\mu_i$ defined in Eq.~\eqref{C10}. To determine the probability
\[
\mathbb P[\text{$i$ belongs to a small cluster $\mathcal C_i$ and $i$ is occupied}],
\]
we proceed as for Eq.\eqref{E12}-\eqref{D41}. Indeed,
\begin{equation}\label{eq:D48}
    \mathbb P[\text{$i$ belongs to a small cluster $\mathcal C_i$ and $i$ is occupied}]
    =
    \mu_i
    \mathbb P\left[
    \text{$i$ belongs to a small cluster $\mathcal C_i$}
    \,\middle|\,
    \text{$i$ is occupied}
    \right],
\end{equation}
where the definition of $\mu_i$ is given in Eq.~\eqref{C10}.
The definition of the conditional probability in Eq.~\eqref{eq:D48} is just
\begin{multline}
    \mathbb P\left[\text{$i$ belongs to a small cluster $\mathcal C_i$} \,\middle|\, \text{$i$ is occupied}
    \right]=\\ 
    \,\,=\Ssum_{s}\sum_{\underline N_i,\{\underline N_{i_\ell}\}_{\ell=1}^{d(i)}}\pi_i\left(S_i=s\middle|\,\underline N_i,\{\underline N_{i_\ell}\}_{\ell=1}^{d(i)}\right)P\left(\underline N_i,\{\underline N_{i_\ell}\}_{\ell=1}^{d(i)}\,\middle|\,a(\underline{N}_i)=1 \right)\\
    \quad=\Ssum_{\{s_\ell\}_{\ell=1}^{d(i)}}\sum_{\underline N_i,\{\underline N_{i_\ell}\}_{\ell=1}^{d(i)}}\pi^{\partial i}\left(\{S^{i_\ell\to i}=s_\ell\}_{\ell=1}^{d(i)}\,\middle|\,\underline N_i,\{\underline N_{i_\ell}\}_{\ell=1}^{d(i)}\right)P\left(\underline N_i,\{\underline N_{i_\ell}\}_{\ell=1}^{d(i)}\,\middle|\,a(\underline{N}_i)=1 \right),
\end{multline}
where $\pi_i$ is the conditional probability that cluster $\mathcal C_i$ is small, conditioned on the neighbors trajectories. The second identity follows  from Eq.~\eqref{E6}, where $\pi^{\partial i}(\cdot | \underline{N}_i,\{\underline{N}_{i_\ell}\}_{\ell=1}^{d(i)})$ is the conditional joint probability for the cluster sizes $\{\mathcal C^{i_{\ell}\to i}\}_{\ell=1}^{d(i)}$. The factorization in Eq.~\eqref{E8} applies also (for the same reason) to $\pi^{\partial i}$, and thus

\begin{equation}
    \mathbb P\left[
    \text{$i$ belongs to a small cluster $\mathcal C_i$}
    \,\middle|\,
    \text{$i$ is occupied}
    \right]
    =
    \sum_{\underline N_i,\{\underline N_{i_\ell}\}_{\ell=1}^{d(i)}}
    \prod_{\ell=1}^{d(i)}
    \left(\sum_s \pi^{i_{\ell}\to i}(s)\right)
    P\left(
    \underline N_i,\{\underline N_{i_\ell}\}_{\ell=1}^{d(i)}
    \,\middle|\,
    a(\underline{N}_i)=1
    \right).
\end{equation}
Thus, all in all, the expression of $\phi_{LC}$ is
\begin{align}
    \phi_{LC}
    &=
    \frac{1}{S}\sum_i
    \mu_i
    \Biggl[
    1
    -
    \sum_{\underline N_i,\{\underline N_{i_\ell}\}_{\ell=1}^{d(i)}}
    \prod_{\ell=1}^{d(i)}
    H^{i_\ell\to i}_{\underline N_{i_\ell},\underline N_i}(1)
    P\left(
    \underline N_i,\{\underline N_{i_\ell}\}_{\ell=1}^{d(i)}
    \,\middle|\,
    a(\underline{N}_i)=1
    \right)
    \Biggr], \label{E24}
\end{align}
where $ P\left(
    \underline N_i,\{\underline N_{i_\ell}\}_{\ell=1}^{d(i)}
    \,\middle|\,
    a(\underline{N}_i)=1
    \right)$ can be obtained, in the same fashion as Eq.~\eqref{E19A} from BDCM, as 
\begin{align}
    P\left(
    \underline N_i,\{\underline N_{i_\ell}\}_{\ell=1}^{d(i)}
    \,\middle|\,
   a(\underline{N}_i)=1
    \right)
    &=
    \frac{
    a(\underline N_i)
    \mathcal A_i\left(\underline N_i,\{\underline N_{i_\ell}\}_{\ell=1}^{d(i)}\right)
    \prod_{\ell=1}^{d(i)}
    \chi^{i_\ell\to i}\left(\underline N_{i_\ell},\underline N_i\right)
    }{
    \sum_{\underline M_i,\{\underline M_{i_\ell}\}_{\ell=1}^{d(i)}}
    a(\underline M_i)
    \mathcal A_i\left(\underline M_i,\{\underline M_{i_\ell}\}_{\ell=1}^{d(i)}\right)
    \prod_{\ell=1}^{d(i)}
    \chi^{i_\ell\to i}\left(\underline M_{i_\ell},\underline M_i\right)
    }.\label{D52}
\end{align}

To obtain the expression for $\phi_{LC}$ in the uniform case reported in Eq.~\eqref{phi_LC_main}, it is sufficient to use again the uniformity over sites and express $\mu_i$ (which is the same for all $i$), together with $P(\{\underline{M}_{\ell}\}_{\ell=1}^{d}, \underline{N}\mid a(\underline{N})=1)$, in terms of the uniform messages, as prescribed in Eq.~\eqref{C10} and Eq.~\eqref{D52}.

\textbf{Finding back independent site percolation from our general recursion.} We show that our general equation Eq.~\eqref{app:D45} reduces to simple site percolation, once correctly evaluated on that process. In the simplest case of site percolation a site is ``active" with a certain probability $p$, independent of the other sites. Thus, in the case of site percolation, the site states are binary. This means that we go from trajectories  $\underline{N}_i$  to states $x\in\{0,1\}$. Similarly, the probability distribution $P(\{x_{u}\}_{u\in\partial i_{\ell}\setminus i}|x_{i_{\ell}}, x_i)$, which corresponds to $P\left(\{\underline N_u\}_{u\in\partial i_{\ell}\setminus i} \,\middle|\, \underline N_{i_\ell},\underline N_i\right)$ in Eq.~\eqref{app:D45}, is just
\begin{equation}
  P(\{x_{u}\}_{u\in\partial i_{\ell}\setminus i}|x_{i_{\ell}}, x_i)=\prod_{u\in \partial i_{\ell}\setminus i}\delta(x_u=1)p+(1-p)\delta(x_u=0).
\end{equation}
The goal is to study the percolation of the $1$ states. We have that Eq.~\eqref{app:D45} specialized to site percolation simplifies to 
\begin{equation}
    H_{x}^{i_{\ell}\to i}(z)=\indi[x=0]+z\indi[x=1]\prod_{u\in\partial i_{\ell}\setminus i}(1-p)H_{0}^{u\to_{i_{\ell}}}+pH_{1}^{u\to_{i_{\ell}}}
\end{equation}
were we lost, due to the independence, the conditioning over the site of the neighbor. 
We then define
$
    \tilde H^{i_{\ell}\to i}(z)
    =
    (1-p)H^{i_{\ell} \to i}_{0}
    +
    pH^{i_{\ell} \to i}_{1}.
$
This is the generating function associated with the unconditional probability
$\pi_{i_{\ell}}(s)$ that $i_{\ell}$ belongs to a small cluster of size $s$ (with no occupation conditioning).
Thus, $\tilde H^{i_{\ell}\to i}$ describes the cluster-size distribution
without conditioning on whether $i_{\ell}$ is occupied.
With this definition, we obtain the following recursion for $\tilde H$.\begin{equation}
    \tilde H^{i_{\ell}\to i}(z)=1-p+pz\prod_{u\in\partial i_{\ell}\setminus i}\tilde H^{u\to i_{\ell}}(z)
\end{equation}
which is exactly the recursion that the generating function of $\pi_i(x)$ must satisfy in site percolation, as detailed in \cite{Site_percolation} (one can do exactly the same argument for bond percolation and find the recursion reported in \cite{PercolationBP}).

\textbf{Simplifying the general percolation recursion with an approximation.} The approach we developed  in the previous section is asymptotically exact, but  solving the recursion in Eq.~\eqref{app:D45} may still be complicated. Thus, we introduce an approximation that significantly simplifies the problem and allows to get a better intuition in the percolation process. Indeed, instead of considering the probability distributions conditioned on the full dynamical trajectory, we can consider the probability distribution conditioned on occupation/vacancy at dynamical convergence. We will then be able to make some approximation that will allow us to obtain a simplified expression for the $\phi_{LC}$ fraction. We start by defining the generating function
\begin{equation}
    \tilde{H}^{i_{\ell}\to i}(z)=\Ssum_s \pi^{i_{\ell}\to i}(\{S^{i_{\ell}\to i}=s\}|a(\underline{N}_i)=1)z^s
\end{equation}
We can now use again the Bayes theorem to write 
\begin{equation}
    \tilde{H}^{i_{\ell}\to i}(z)= P\left(a(\underline{N}_{i_\ell})=0\middle|a(\underline{N}_i)=1\right)+P\left(a(\underline{N}_{i_\ell})=1\middle|a(\underline{N}_i)=1\right)\Ssum_s\pi^{i_{\ell}\to i}(\{S^{i_{\ell}\to i}=s\}|a(\underline{N}_i)=1,a(\underline{N}_{i_\ell})=1)z^s.
\end{equation}
We can then recognize the definition of $\mu_{i_{\ell}|i}$ from Eq.~\eqref{C10}, while from Eq.~\eqref{E7} we derive the following:

\begin{equation}
\begin{aligned}
&\sum_s z^s\,
\pi^{i_{\ell}\to i}
\left(
    \{S^{i_{\ell}\to i}=s\}
    \,\middle|\,
    a(\underline{N}_i)=1,\,
    a(\underline{N}_{i_\ell})=1
\right)
\\
&\qquad =
z
\sum_{\{s_u\}_{u\in\partial i_{\ell}\setminus i}}
\pi^{\partial i_{\ell}\setminus i\to i_{\ell}}
\left(
    \{S^{u\to i_{\ell}}=s_u\}_{u\in\partial i_{\ell}\setminus i}
    \,\middle|\,
    a(\underline{N}_i)=1,\,
    a(\underline{N}_{i_\ell})=1
\right)
\prod_{u\in\partial i_{\ell}\setminus i} z^{s_u}.
\end{aligned}
\end{equation}

All in all, this allows to write
\begin{equation}
    \tilde{H}^{i_{\ell}\to i}(z)=1-\mu_{i_{\ell}|i}+\mu_{i_{\ell}|i}z\hspace{-4mm}\Ssum_{\{s_k\}_{k=1}^{d(i)-1}}\hspace{-3mm}\pi^{\partial i_{\ell}\setminus i\to i_{\ell}}\left(\hspace{-1mm}\{S^{u\to i_{\ell}}=s_k\}_{\substack{
u\in\partial i_{\ell}\setminus i\\
k\in\{1,\dots,d(i_{\ell})-1\}
}} \,\middle|a(\underline{N}_i)=1,a(\underline{N}_{i_\ell})=1\hspace{-1mm}\right)\hspace{-1mm}\prod_{k=1}^{d(i)-1}\hspace{-2mm}z^{s_k}.\end{equation}
As of now, everything is exact. We can now introduce the following approximation: we assume that the joint probability $\pi^{\partial i_{\ell}\setminus i\to i_{\ell}}(\cdot |a(\underline{N}_i)=1,a(\underline{N}_{i_\ell})=1)$, conditioned just on the final state occupancy, is factorized and that the first condition over $a(\underline{N}_i)=1$ does not matter.  We stress that this is an approximation, as the conditional probability factorizes only when conditioning over the full trajectories. Thus, we can  write 
\begin{equation}\label{app:D56}
    \pi^{\partial i_{\ell}\setminus i\to i_{\ell}}(\{S^{u\to i_{\ell}}\}_{u\in\partial i_{\ell}\setminus i} |a(\underline{N}_i)=1,a(\underline{N}_{i_\ell})=1)\approx \prod_{u\in\partial i_{\ell}\setminus i} \pi^{u\to i_{\ell}}(S^{u\to i_{\ell}}|a(\underline{N}_{i_{\ell}})=1).
\end{equation}
Then, under this approximation, we obtain that 
\begin{equation}\label{simplerecursion}
    \tilde{H}^{i_{\ell}\to i}(z)\approx 1-\mu_{i_{\ell}|i}+z\mu_{i_{\ell}|i}\prod_{u\in\partial i_{\ell}\setminus i}\tilde{H}^{u\to i_{\ell}}(z).
\end{equation}
We can notice that this is just a site percolation process with non uniform, independent occupation probability $\mu_{i_{\ell}|i}$. Thus, the approximation in Eq.~\eqref{app:D56} maps back the general percolation process to a site percolation process.

Furthermore, for a $d$-regular graph, one can obtain an approximate criterion for the onset of percolation by looking at when the trivial fixed point $\tilde H =1$ of \eqref{simplerecursion} loses stability (as discussed in page 3 in \cite{PercolationBP}). This happens exactly at $\mu^{\rm conditional}\approx\frac{1}{d-1}$ ($\mu_{j|i}=\mu^{\rm conditional}$ for all $i,j$ for a $d$-regular graph). 
From the recursion in Eq.~\eqref{simplerecursion} one can also obtain an approximated formula for $\phi_{LC}$ (similar to the one in \cite{PercolationBP})
\begin{equation}
    \phi_{LC}\approx\frac{1}{S}\sum_{i}\Bigl(\mu_i-\mu_i\prod_{\ell=1}^{d(i)}\tilde{H}^{i_{\ell}\to i}(1)\Bigr).
\end{equation}
To assess how good this approximation is, we compare the $\phi_{LC}$ fraction computed with the approximation and the exact recursion. We find that the difference between approximated and exact value is of order $10^{-2}$ for almost all  $\alpha$ and $d$ (for $d=5$ the difference is more pronounced), the comparison is done in Figure~\ref{all_d_K_2} for $K=2$.

\section{Numerical methods to solve BDCM on $d$--regular graphs}\label{FFT}

We now detail the numerical procedure used to solve efficiently the BDCM equations Eq. \eqref{C11}, for $d$--regular graphs and uniform couplings. To make computations faster, we introduce the concept of a dynamical programming matrix.  We note that the constraint matrix $\mathcal{A}(\underline{N}, [\underline{\mathbf{y}}]_{d-1} \cup \underline{M}\})$ does not  effectively depend on the specific trajectory of all the neighbors, but instead on the \emph{cumulant trajectory}: \[\underline{\tilde{k}}=\sum_{\underline{y}'\in [\underline{\mathbf{y}}]_{d-1} \cup \underline{M} }\underline{y}'\]
To understand this, it is sufficient to inspect the dynamical rule in  Eq.~\eqref{discrete_glv}, and note that the next state depends only on $\sum_{j\in\partial i}y_j$. We insert the definition  of the cumulant trajectory with an indicator function,
where we have singled-out the trajectory $\underline{M}$. The update becomes
\begin{equation}
    \chi_{\underline{N},\underline{M}}=\sum_{\underline{N},[\underline{\mathbf{y}}]_{d-1}}{\mathcal{A}(\underline{N}, [\underline{\mathbf{y}}]_{d-1} \cup \underline{M})\prod_{\underline{y}\in [\underline{\mathbf{y}}]_{d-1}}\chi_{\underline{y},\underline{N}}^{\rightarrow}}=\sum_{\underline{N},\underline{k}}{\mathcal{A}(\underline{N}, \underline{M},\underline{k})\sum_{[\underline{\mathbf{y}}]_{d-1}}\indi\Bigl[\underline{k}-\sum_{\underline{y}'\in [\underline{\mathbf{y}}]_{d-1} }\underline{y}'\Bigr]\prod_{\underline{y}\in [\underline{\mathbf{y}}]_{d-1}}\chi_{\underline{y},\underline{N}}^{\rightarrow}}.
\end{equation}
We now define the dynamical programming matrix \[
W_{d-1}(\underline{k},\underline{N})=\sum_{[\underline{\mathbf{y}}]_{d-1}}\indi\Bigl[\underline{k}-\sum_{\underline{y}'\in [\underline{\mathbf{y}}]_{d-1} }\underline{y}'\Bigr]\prod_{\underline{y}\in [\underline{\mathbf{y}}]_{d-1}}\chi_{\underline{y},\underline{N}}^{\rightarrow}
\]
which follows the convolution
\[
W_{d-1}(\underline{k},\underline{N})=\sum_{\underline{x}}W_{d-2}(\underline{k}-\underline{x},\underline{N})\chi^{\rightarrow}_{\underline{x},\underline{N}},
\]
where the sum over $\underline{x}$ is a sum over all the possible sequences of lengths $p+c$ in which each entry can take values from $0$ to $K$ (extremes included). Calling $\mathcal{F_{D}}$ the (multidimensional) discrete Fourier transform over the first $p+c$ arguments (i.e. this transform leaves the $\underline{N}$ unaltered) it follows
\[
W_{d-1}(\underline{k},\underline{N})=\mathcal{F_{D}}^{-1}\Bigg[\Bigl(\mathcal{F_{D}}[\chi^{\rightarrow}_{\underline{x},\underline{N}}]\Bigr)^{d-1}\Bigg](\underline{k},\underline{N})
\]
The complexity of solving the BP equations of BDCM is effectively of order $\mathcal{O}(dK^{p+c})$. The implementation of the code is done in PyTorch \cite{torch}. The drawback of this approach is that it requires storing matrices which are exponentially large in $d$. 
\subsection{Computing the structure functions $\eta(\ell)$ with FFT  }\label{ETALFTT}
\label{etal}
To compute this observable, we introduce the following edge localized observable
\begin{equation}
    \tilde{\eta_l}(\mathbf{N}_{p+1})=\frac{1}{S}\sum_{i=1}^S\indi[N_i^{p+1}>0]\indi\Bigl[\sum_{j\in\partial i}\indi[N_j^{p+1}>0]=l\Bigr]=\frac{1}{S}\sum_{i}^S\tilde{\eta}_l^{(i)}(N_i^{p+1},\{N_j^{p+1}\}_{j\in\partial i}).
\end{equation}
Remember that if a site is occupied at time $p+1$ it will be occupied over the full attractor. $\tilde{\eta}_l^{(i)}(N_i^{p+1},\{N_j^{p+1}\}_{j\in\partial i})$ is an indicator function that equals $1$ if site $i$ is occupied and connected to $l$ occupied neighbors, and $0$ otherwise. Thus, summing over all the $\tilde{\eta}_l^{(i)}(N_i^{p+1},\{N_j^{p+1}\}_{j\in\partial i})$ corresponds to counting the number of occupied sites with $l$ occupied neighbors. 
Now, the degree distribution $\eta(\ell)$ is given by
\begin{equation}
    \eta(l)=\frac{\langle \tilde{\eta_l}(\mathbf{N}_{p+1})\rangle}{1-\langle \rho_0(\mathbf{N})\rangle},
\end{equation}
where we normalize by the fraction of occupied sites, in order to have $\sum_{l=0}^{d}\eta(l)=1$. We consider again the case of a $d$--regular graph (uniform interactions), so all sites are equivalent and thus
\begin{equation}
    \langle \tilde{\eta_l}(\mathbf{N}_{p+1})\rangle = \langle \tilde{\eta_l}^{(i)}(N_i,\{N_j\}_{j\in\partial i})\rangle.
\end{equation}
As for any other observables we have that
\begin{equation}\label{C4}
     \langle \tilde{\eta_l}(\mathbf{N}_{p+1})\rangle = \frac{\sum_{\underline{x},\{\underline{{y}}_i\}_{i=1}^{d}}\tilde\eta_{l}^{(\cdot)}(x^{p+1}, \{y_i^{p+1}\}_{i=1}^{d})\mathcal{A}(\underline{x},\{\underline{{y}_i}\}_{i=1}^{d})\prod_{\underline{y}\in\{\underline{{y}}_i\}_{i=1}^{d}}\chi_{\underline{x},\underline{y}}}{Z_{\text{fac}}}.
\end{equation}
The difficulty in evaluating this expression is that the dynamical programming method, in the previously presented form, is not applicable anymore, meaning that computing this expression directly would not be possible (the number of operations would be $O(K^{d(p+c)})$). However, there is a useful simplification that arises for the observable in question. Let's focus on the numerator of  (\ref{C4}), ignoring the sum over the $\underline{x}$. We can  define the following:
\begin{equation}\label{C5}
    U_l(\underline{x})\equiv\sum_{\{\underline{\mathbf{y}}\}_{i=1}^{d}} \indi\Bigl[\sum_{i=1}^{d}\indi[y_i^{p+1}>0]=l\Bigr]\mathcal{A}(\underline{x},\{\underline{{y}}_i\}_{i=1}^{d})\prod_{i=1}^{d}\chi_{\underline{x},\underline{y_i}}.  
\end{equation}
Let $b_i$ be a binary variable defined as $b_i=\indi[y_i^{p+1}>0]$. Notice furthermore that, for a fixed set of $\{y_i\}_{i=1}^{d}$, \[
\sum_{\{b_i\}_{i=1}^{d}\in\{0,1\}^{d}}\prod_{i=1}^{d}\indi\Bigl[b_i=\indi\{y_i^{p+1}>0\}\Bigr] =1.
\]
After multiplying (\ref{C5}) by this identity and exchanging the order of summation, we get that 
\begin{equation}
   U_l(\underline{x})= \sum_{\{b_i\}_{i=1}^{d}\in\{0,1\}^{d}}\indi\Bigl[\sum_{i=1}^{d}b_i=l\Bigr]\sum_{\{\underline{y}_i\}_{i=1}^{d}}\mathcal{A}(\underline{x},\{\underline{{y}}_i\}_{i=1}^{d})\prod_{i=1}^{d}\indi\Bigl[b_i=\indi\{y_i^{p+1}>0\}\Bigr]\chi_{\underline{x},\underline{y}_i}.
\end{equation}
Now we define two new auxiliary functions, $v_0(\underline{x},\underline{y}_i)=\indi\Bigl[0=\indi\{y_i^{p+1}>0\}\Bigr]\chi_{\underline{x},\underline{y}_i}$ and $v_1(\underline{x},\underline{y}_i)=\indi\Bigl[1=\indi\{y_i^{p+1}>0\}\Bigr]\chi_{\underline{x},\underline{y}_i}$. Thus $U_l(\underline{x})$ is a convolution of $v_0$ and $v_1$, and we obtain
\begin{equation}
\label{C7}
    U_l(\underline{x})= \sum_{\{b_i\}_{i=1}^{d}\in\{0,1\}^{d}}\indi\Bigl[\sum_{i=1}^{d}b_i=l\Bigr]\sum_{\underline{k}}\mathcal{A}(\underline{x},\underline{k})\mathcal{F}^{-1}\Bigl[\mathcal{F}[v_0]^{d-l}\mathcal{F}[v_1]^{l}\Bigr](\underline{x},\underline{k}).
\end{equation}
So all in all the average value of $\tilde{\eta_l}(\mathbf{N}_{p+1})$ reads:
\begin{equation}
\label{C8}
     \langle \tilde{\eta_l}(\mathbf{N}_{p+1})\rangle = \binom{d}{l}\frac{\sum_{\underline{x},\underline{k}}\indi[x^{p+1}>0]\sum_{\underline{k}}\mathcal{A}(\underline{x},\underline{k})\mathcal{F}^{-1}\Bigl[\mathcal{F}[v_0]^{d-l}\mathcal{F}[v_1]^{l}\Bigr](\underline{x},\underline{k})}{Z_{\text{fac}}},
\end{equation}
where the binomial factor comes from the sum over $\{b_i\}_{i=1}^{d}$ in equation (\ref{C7}). This means that we are able to compute $\eta_l$ efficiently, as the complexity of computing (\ref{C8}) is of order $\mathcal{O}(dK^{p+c})$, compared to $\mathcal{O}{(K^{d(p+c)}})$ for the ``row" expression of $\eta_l$ in \ref{C4}. 

\subsection{General properties about the percolation fixed point recursion and discussion about the initialization}
We start by firstly defining the shorthand $q^{i_\ell\to i}_{\underline N_{i_\ell},\underline N_i}=H^{i_\ell\to i}_{\underline N_{i_\ell},\underline N_i}(1)$ and to write the percolation recursion in Eq.~\eqref{E19}, specialized to the BDCM case, as 
\begin{align}
     q^{i_\ell\to i}_{\underline N_{i_\ell},\underline N_i}
     &=
     \begin{cases}
    1,
     & \text{if } A_{i_\ell}=0,\\[1mm]
     \displaystyle
     \frac{
     \sum_{\{\underline N_u\}_{u\in\partial i_\ell\setminus i}}
     \mathcal A_{i_\ell}\left(
     \underline N_{i_\ell},\underline N_i,
     \{\underline N_u\}_{u\in\partial i_\ell\setminus i}
     \right)
     \prod_{u\in\partial i_\ell\setminus i}
     q^{u\to i_\ell}_{\underline N_u,\underline N_{i_\ell}}
     \chi^{u\to i_\ell}\left(\underline N_u,\underline N_{i_\ell}\right)
     }{
     \sum_{\{\underline M_u\}_{u\in\partial i_\ell\setminus i}}
     \mathcal A_{i_\ell}\left(
     \underline N_{i_\ell},\underline N_i,
     \{\underline M_u\}_{u\in\partial i_\ell\setminus i}
     \right)
     \prod_{u\in\partial i_\ell\setminus i}
     \chi^{u\to i_\ell}\left(\underline M_u,\underline N_{i_\ell}\right)
     },
     & \text{if } A_{i_\ell}=1.
     \end{cases}
     \label{fixedpointrecursion}
 \end{align}

The expression for $q$ in Eq.~\eqref{fixedpointrecursion} has interesting properties that warrant some discussion. First, we note that the equation always admits the trivial solution
\[
q^{{\rm trivial }i_\ell\to i}_{\underline N_{i_\ell},\underline N_i}=1.
\]
This corresponds to the non-percolating solution. It is the physical solution in the non-percolating phase, while in the percolative phase, there will exist another solution $q^*$ with $q^*< q^{\rm trivial}$ (the $<$ is to be interpreted component-wise). This is similar to what happens in independent percolation systems, as discussed in~\cite{PercolationBP}, although in our case the percolation is indirect and comes from the dynamics.

Another important observation is about the monotonicity of the update map, as we shall now explain. Let $\mathcal T^{i_\ell\to i}$ denote the right-hand side of Eq.~\eqref{fixedpointrecursion}. To solve Eq.~\eqref{fixedpointrecursion}, one initializes the $q$'s and iterates
\begin{equation}\label{E27}
    q^{i_\ell\to i}_{\underline N_{i_\ell},\underline N_i}
    \leftarrow
    \mathcal T^{i_\ell\to i}
    \left[
    \{q^{u\to i_\ell}\}_{u\in\partial i_\ell\setminus i}
    \right]
\end{equation}
until convergence. The physical initialization is
\begin{equation}
   q^{\rm physical,\,i_\ell\to i}_{0;\underline N_{i_\ell},\underline N_i}
   =
   \begin{cases}
      1, & \text{if } A_{i_\ell}=0,\\
      0, & \text{if } A_{i_\ell}=1.
  \end{cases}
\end{equation}
Indeed, if a site is unoccupied, the cavity cluster rooted at that site has size zero and is therefore small with probability one. Conversely, initializing $q=0$ on occupied trajectories corresponds to assuming that every occupied site may belong to the giant component. In other words, this is the ``smallest possible initialization" and as will become apparent below, it is the one that will allow us to find the $q^*$ percolating solution, if it exists (if it does not exist, this initialization will just converge to the trivial solution).

With this initialization, the following sequence is monotone:
\begin{equation}\label{E29}
    \mathcal T^n[q^{\rm physical}_0]
    \geq
    \mathcal T^{n-1}[q^{\rm physical}_0],
\end{equation}
where $\geq$ is to be interpreted component-wise (this follows from the fact that the message and $\chi$ are positive, thus component-wise $q\geq q'$ $\mathcal{T}[q]\geq\mathcal{T}[q']$ and furthermore $\mathcal{T}[q_0^{\rm physical}]\geq q_0^{\rm physical}$). Therefore, the iteration converges to the smallest fixed point of Eq.~\eqref{fixedpointrecursion}. If the system is in the non-percolating phase, this smallest fixed point is the trivial solution $q=1$. If the system is in the percolating phase, the smallest fixed point satisfies $q^{i\to j}_{\underline{N}_i,\underline{N}_j}<1$ for some $i,j$ (an extensive fraction).

\section{Supporting results for the BDCM computation and the fully connected limit}\label{Additonal_BDCM}
\subsection{$\rho_0$, $\phi_{LC}$ $\eta_l$ and thresholds at $K=3$ and higher $K$}\label{K=3}
\begin{figure*}[t]
\centering

\hfill
\begin{minipage}{0.66\textwidth}
    \centering
    \includegraphics{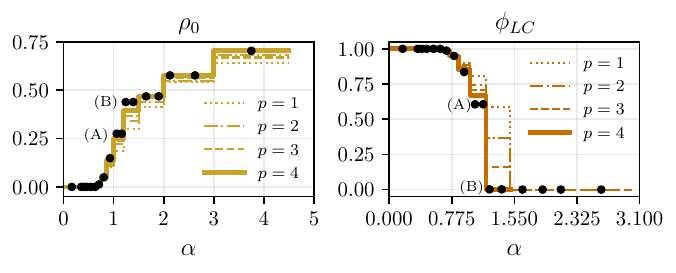} 
\end{minipage}
\hfill
\begin{minipage}{0.32\textwidth}
    \centering
    \renewcommand{\arraystretch}{1.2}
    \setlength{\tabcolsep}{6pt}
    \begin{tabular}{r @{\hspace{10pt}} c c @{\hspace{18pt}} c c}
    \hline
    \multicolumn{5}{c}{BDCM entropy and $\phi_{LC}$}\\
    \multicolumn{5}{c}{for hard-to-converge $\alpha$}\\
    \hline
    & \multicolumn{2}{c}{$\alpha=1.16$ (A)} & \multicolumn{2}{c}{$\alpha=1.242$ (B)}\\
    $p$ & $\frac{\Phi_{(p/c)}}{\log K}$ & $\phi_{LC}$ 
        & $\frac{\Phi_{(p/c)}}{\log K}$ & $\phi_{LC}$ \\
    \hline
    1 & 0.8855 & 0.8047 & 0.7696 & 0.5895 \\
    2 & 0.9529 & 0.7412 & 0.8796 & 0.3636 \\
    3 & 0.9743 & 0.7069 & 0.9264 & 0.1592 \\
    4 & 0.9848 & 0.6710 & 0.9539 & 0.0000 \\
    \hline
    \end{tabular}
\end{minipage}
\hfill

\vspace{0.35cm}

\begin{minipage}{\textwidth}
    \centering
    \includegraphics{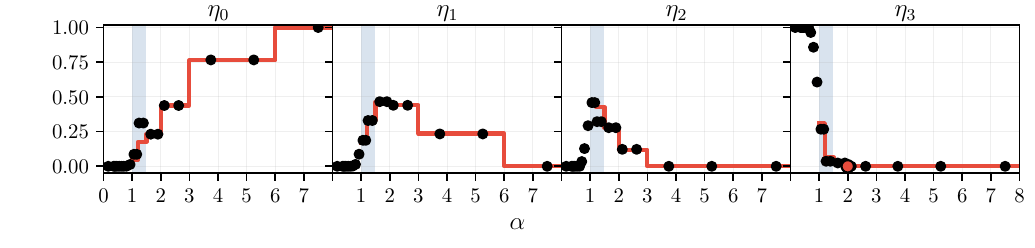}
\end{minipage}

\caption{\textbf{BDCM captures typical dynamics and structure for $K=3$, with slower entropy convergence near percolation.}
\textit{(Top)} Extinction fraction $\rho_0$ and largest cluster fraction $\phi_{LC}$ as functions of $\alpha$, computed from BDCM for the dominant $(p/c)$--attractor ($c=2$) and multiple values of $p$ ($\phi_{LC}$ has been  computed with the exact expression, but for $p=4$ the computation was carried out only slightly beyond the threshold of percolation, given that after that threshold it is just zero). Black dots denote numerical simulations on $d$--regular graphs ($S=10^4$), showing excellent agreement. Agreement is essentially perfect for $\alpha \leq 1$ and $\alpha \geq 3/2$, where at $p=4$ the entropy coverage exceeds $99\%$. In the intermediate regime $1 \leq \alpha \leq 3/2$, convergence is slower; the table reports representative values of $\Phi/\log(K)$ and $\phi_{LC}$ (for $\alpha$ values corresponding to the black dots), showing that entropy coverage remains above $95\%$.
\textit{(Bottom)} Structure functions $\eta_l$ as a function of $\alpha$ for the same attractor ($p=4$, $c=2$). The shaded region marks $1 \leq \alpha \leq 3/2$, where slower entropy convergence leads to small discrepancies with simulations. The linear community transition occurs at $\alpha_{\rm lin}=2$ (the red marker). Note that here we do not report the theory prediction for small $\alpha$, given the computational cost of computing $\eta_l$ and the fact that we use it only to extract information of $\alpha_{\rm lin}$}

\label{appendix_K_3}
\end{figure*}

We plot in Figure~\ref{appendix_K_3} $\phi_{LC}$ and $\rho_0$ obtained from BDCM as functions of $\alpha$ for $K=3$ (at $d=3$), reproducing Figure~\ref{fig2} from the main text and comparing with empirical simulations (in particular $\phi_{LC}$ has been computed with the asymptotically exact expression). Also in this case the dominant attractor is a $2$-cycle (the entropy for the $(p/c)$-backtracking attractor at $c=2$ is larger than the one at $c=1$, see Figure~\ref{c1vsc2}). The general trend of the observables for $K=3$ is virtually identical to that for $K=2$. However, while for $K=2$ we reach more than $99\%$ entropy coverage for all $\alpha$ (at the largest $p$ considered), for $K=3$ and $1 \leq \alpha \leq 3/2$ the maximal entropy coverage reaches between $95\%$ and $98\%$ (while for other values of $\alpha$ it remains above $99\%$).

Thus, the observables computed with BDCM, for $1 \leq \alpha \leq 3/2$, do not exactly match those obtained from empirical simulations. Nevertheless, we are still able to infer both the percolation and extinction thresholds. For the latter, we note that the first value of $\alpha$ for which $\rho_0 > 0$ is associated with an entropy that is already around $99\%$ of the total, ensuring that this is the critical threshold for typical initializations. Thus, $\alpha_{\rm ext}=2/3$ (already at $p=2$ at $\alpha_{\rm ext}=2/3$ the entropy is around $99\%$).
For the percolation threshold, we observe that $\phi_{LC}$ is monotonically decreasing in $p$ and satisfies $\phi_{LC} \geq 0$. This allows us to argue that the percolation threshold is at $\alpha_{\rm perc} = 6/5$: for $\alpha \geq 6/5$, we find $\phi_{LC} = 0$ already at $p=4$, while for $\alpha \leq 6/5$, $\phi_{LC}$ remains significantly non-zero, and in this regime the entropy coverage is already around $98\%$. 

Regarding the linear community $\alpha_{\rm lin}$ critical threshold (still for $d=3$), we show the $\eta_l$ as a function of $\alpha$, and note that $\eta_3$ goes to zero for $\alpha_{\rm lin}=2$, and for this value of interaction strength the entropy turns out to be above $99\%$ of the full entropy (at the maximal $p=4$ considered).

We show in the Table~\ref{Diagram_K=3} the threshold for multiple $d$ also for $K=3$ (but at higher finite $d$ we compute analytically only the extinction threshold), performing numerical simulations and carrying out our analytical procedure when possible. 
For higher $K$ the threshold, which we again show in  Table~\ref{Diagram_K=3}, are computed from numerical simulations, and thus have an errorbar associated to them. Indeed, in those cases, BDCM quickly becomes computationally intractable, and only $p=1$ can be considered. This does not allow us to reach sufficiently high entropy to obtain reliable estimates of the observables for all the $\alpha$ values.

\begin{table*}[t]
\centering
\scriptsize
\setlength{\tabcolsep}{2.2pt}
\renewcommand{\arraystretch}{1.35}
\resizebox{\textwidth}{!}{%
\begin{tabular}{c|ccc|ccc|ccc|ccc|ccc|ccc|}
\hline
 & \multicolumn{3}{c|}{$K=2$}
 & \multicolumn{3}{c|}{$K=3$}
 & \multicolumn{3}{c|}{$K=4$}
 & \multicolumn{3}{c|}{$K=30$}
 & \multicolumn{3}{c|}{$K=40$}
 & \multicolumn{3}{c}{$K=50$} \\
\cline{2-19}
$d$
& $\alpha_{\mathrm{lin}}$ & $\alpha_{\mathrm{ext}}$ & $\alpha_{\mathrm{per}}$
& $\alpha_{\mathrm{lin}}$ & $\alpha_{\mathrm{ext}}$ & $\alpha_{\mathrm{per}}$
& $\alpha_{\mathrm{lin}}$ & $\alpha_{\mathrm{ext}}$ & $\alpha_{\mathrm{per}}$
& $\alpha_{\mathrm{lin}}$ & $\alpha_{\mathrm{ext}}$ & $\alpha_{\mathrm{per}}$
& $\alpha_{\mathrm{lin}}$ & $\alpha_{\mathrm{ext}}$ & $\alpha_{\mathrm{per}}$
& $\alpha_{\mathrm{lin}}$ & $\alpha_{\mathrm{ext}}$ & $\alpha_{\mathrm{per}}$ \\
\hline

$3$
& \bv{1} & \bv{\frac{1}{2}} & \bv{\frac{3}{4}}
& \bv{2} & \bv{\frac{2}{3}} & \bv{\frac{6}{5}}
& $\frac{9}{4}$ & $\frac{3}{4}$ & \err{1.7}{1}
& \err{1.85}{3} & \err{1.165}{4} & \err{1.37}{1}
& \err{1.76}{4} & \err{1.149}{5} & \err{1.329}{8}
& \err{1.74}{2} & \err{1.137}{3} & \err{1.321}{8}
\\
\hline

$4$
& \bv{\frac{4}{3}} & \bv{\frac{1}{2}} & \bv{\frac{4}{3}}
& $\frac{8}{3}$ & \bv{\frac{2}{3}} & $\frac{8}{5}$
& $3$ & $\frac{3}{4}$ & \err{2.2}{2}
& \err{2.41}{4} & \err{1.278}{6} & \err{1.73}{2}
& \err{2.33}{3} & \err{1.251}{4} & \err{1.70}{1}
& \err{2.23}{3} & \err{1.245}{5} & \err{1.68}{2}
\\
\hline

$5$
& \bv{\frac{5}{3}} & \bv{\frac{1}{2}} & \bv{\frac{5}{4}}
& $\frac{10}{3}$ & \bv{\frac{2}{3}} & \err{2.1}{2}
& $\frac{15}{4}$ & $\frac{3}{4}$ & \err{2.9}{2}
& \err{2.95}{8} & \err{1.375}{6} & \err{2.16}{2}
& \err{2.88}{5} & \err{1.355}{7} & \err{2.11}{2}
& \err{2.80}{3} & \err{1.343}{5} & \err{2.08}{2}
\\
\hline

$6$
& \bv{2} & \bv{\frac{1}{2}} & \bv{\frac{3}{2}}
& $4$ & \bv{\frac{2}{3}} & \err{3.1}{6}
& $\frac{9}{2}$ & \err{0.77}{2} & \err{3.2}{6}
& \err{3.53}{5} & \err{1.45}{1} & \err{2.58}{3}
& \err{3.42}{3} & \err{1.464}{8} & \err{2.53}{2}
& \err{3.34}{9} & \err{1.451}{9} & \err{2.48}{2}
\\
\hline

$7$
& \bv{\frac{7}{3}} & \bv{\frac{1}{2}} & \bv{\frac{7}{4}}
& $\frac{14}{3}$ & \bv{\frac{2}{3}} & $\frac{7}{2}$
& $\frac{21}{4}$ & \err{0.77}{3} & $\frac{21}{5}$
& \err{4.00}{7} & \err{1.50}{1} & \err{2.99}{3}
& \err{3.97}{7} & \err{1.51}{1} & \err{2.93}{2}
& \err{3.88}{8} & \err{1.53}{1} & \err{2.91}{3}
\\
\hline

$\infty$
& -- & \bv{\frac{2}{3}} & --
& -- & \bv{1} & --
& -- & \bv{\frac{6}{5}} & --
& -- & -- & --
& -- & -- & --
& -- & -- & --
\\
\hline

\end{tabular}
}
\caption{
Thresholds $\alpha_{\mathrm{lin}}$, $\alpha_{\mathrm{ext}}$, and $\alpha_{\mathrm{per}}$
for different values of $K$ and $d$. Shaded entries correspond to exact theoretical predictions, non shaded entries to numerical simulations. In the $d=\infty$ limit the percolation and linear  community thresholds are not well defined, and we compute analytically $\alpha_{\rm ext}$ only at small $K$. The results obtained from the BDCM theory are computed with the highest available $p=4$. Number in parentheses  must be interpreted as errorbars over last digit. Non-shaded fraction corresponds to threshold for which all the numerical simulations conducted agree about the threshold position (but for which we do not have the theoretical prediction).
}
\label{Diagram_K=3}
\end{table*}

\subsection{Computing the critical extinction threshold and the atypical threshold at $d=\infty$}\label{High_d_appendix}
\begin{figure}
    \centering
    \includegraphics[width=0.8\textwidth]{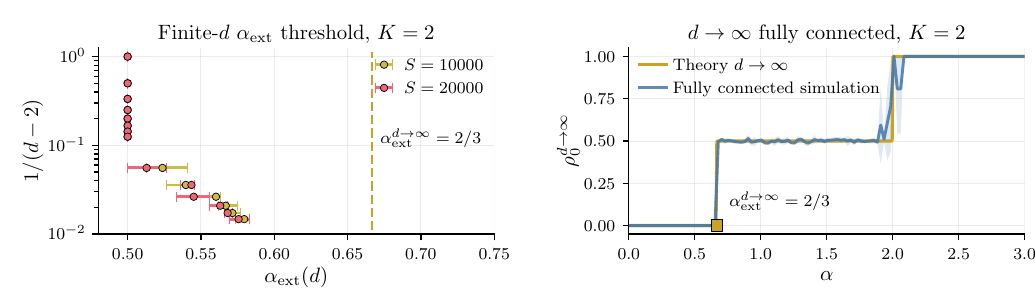}
    \caption{\textbf{The high degree limit of BDCM equation, $K=2$.} \textit{(Left)} Numerical computation of the extinction threshold at high $d$ (over 5 simulations, $S=10000, S=20000$). Notice the drift from $\alpha_{\rm ext}=1/2$ (we report simulations for two sizes to illustrate that this drift is not a finite size effect).  \textit{(Right)} Theoretical prediction of $\rho_0$ (in red, from solving numerically Eq.~\eqref{C24}) in the fully connected limit, alongside a numerical comparison for a fully connected graph (simulation done on $S$ sites). It is possible to see that the extinction threshold (at $d=\infty$) is located at $\alpha_{\rm ext}=2/3$ for $K=2$. }
    \label{high_degree_K_2}
\end{figure}

\begin{figure}
    \centering
    \includegraphics[width=0.8\textwidth]{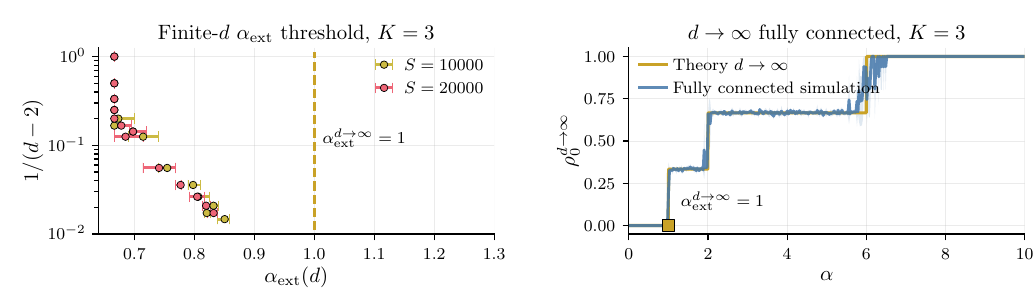}
    \caption{\textbf{The high degree limit of BDCM equation, $K=3$.} See caption of Figure~\ref{high_degree_K_2}. It can be seen that the extinction threshold is at $\alpha_{\rm ext}=1$ in the fully connected limit. Also, in this case, there is a drift from  $\alpha_{\rm  ext}=2/3$ (which is the extinction threshold at small $d$) in the extinction threshold as $d$ is increased.}
    \label{high_degree_K_3}
\end{figure}

We present in Figure~\ref{high_degree_K_2} ($K=2$) and Figure~\ref{high_degree_K_3} ($K=3$) the numerical solutions of Eq.~\eqref{C23}, which allow us to compute the extinction fraction $\rho_0$ (via Eq.~\eqref{C24}) in the fully connected limit ($d=\infty$), and thus determine the corresponding extinction threshold $\alpha_{\rm ext}^{d\to\infty}$. These results support the claim (see Figure~\ref{fig1} and Table~\ref{Diagram_K=3}) that the critical threshold $\alpha_{\rm ext}$ at analytically accessible ``small'' $d$ (i.e., $d\leq 8$) differs from its value in the fully connected limit.

To validate these predictions, we perform numerical simulations (for systems with $S= 1000$ sites) on fully connected graphs, comparing the empirically measured extinction fraction with the theoretical prediction for $\rho_0$ from Eq.~\eqref{C24}. We also show via numerical simulations that, at very large $d$ (beyond the reach of the BDCM computation), the critical extinction threshold drifts away from the constant value observed at small $d\leq 8$.

\begin{figure}
    \centering
    \begin{minipage}{\linewidth}
        \centering
        \includegraphics{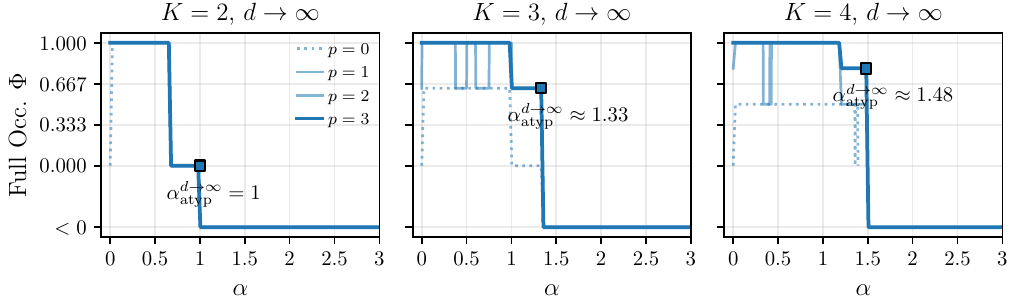}
    \end{minipage}
    \caption{\textbf{Computation of the Extinction-Forbidden entropy at $K=2,3,4$ for  $d\to\infty$} We show the extinction forbidden entropy at  $d\to\infty$, to determine the $\alpha_{\rm atyp}^{d\to\infty}$ threshold. In the $K=4$ plot note that the $p=0$ entropy jumps from $0$ to  $\approx 0.5$ once at $\alpha\approx 1.2$. }
    
    \label{figure_d_infty_atyp}
\end{figure}

We then report in Figure \ref{figure_d_infty_atyp} (still for $K=2,3,4$) the extinction forbidden entropy $\Phi^{\rm Full.Occupied}_{(p/c)}$ in the $d\to\infty$ limit. This is computed through Eq.~\eqref{C25} (with the constraint matrix in Eq.~\eqref{C27}). The value of alpha for which this is negative (in this case the entropy is actually $-\infty$)  is the $\alpha_{\rm atyp}^{d\to \infty}$ threshold. 

\subsection{Is the dominant attractor a cycle or a fixed point?}\label{cyclefixedpoint}
\begin{figure}[!h]
    \centering
    \begin{minipage}{\linewidth}
        \centering
        \includegraphics{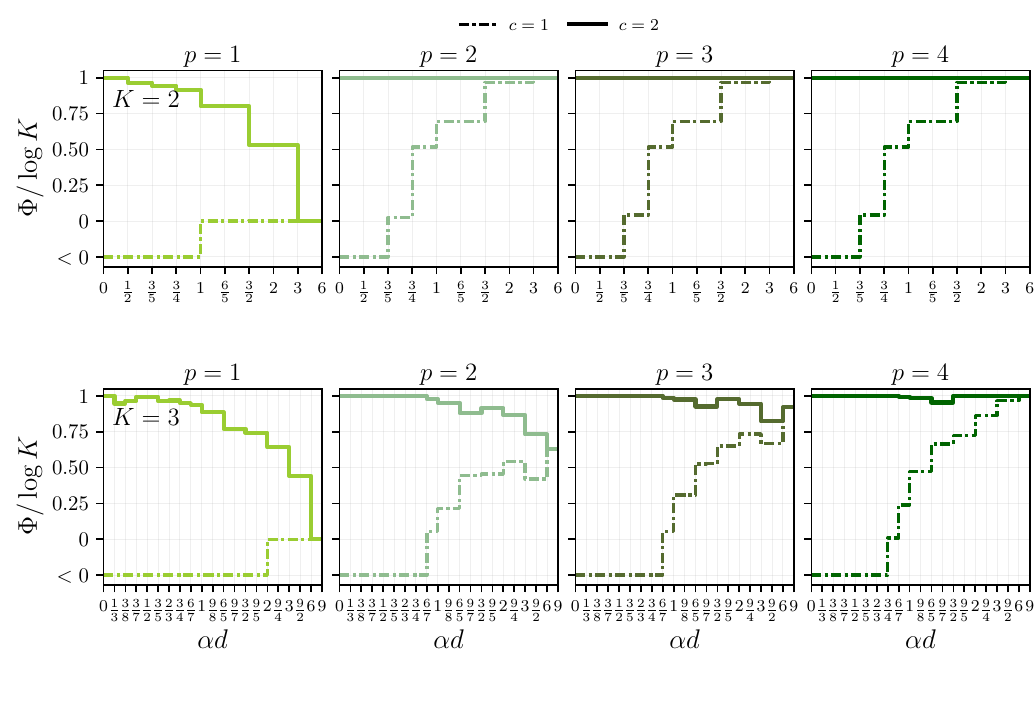}
    \end{minipage}
    \caption{\textbf{Entropy $\Phi_{(p/c)}$ for $c=1$ and $c=2$ at $K=2,3$, for $3$--random-regular graphs.}
     The entropy of $c=2$ attractors is always larger or equal than the entropy of $c=1$ fixed points (and is strictly larger at ``small" $\alpha$). At very large $\alpha$, independent set fixed points become dominant and the two entropies are equal.}
    \label{c1vsc2}
\end{figure}

We show in Figure~\ref{c1vsc2}, for $K=2$ and $K=3$, the comparison of the entropies of the $(p/c)$-backtracking attractors with $c=2$ and $c=1$. This shows that the dominant attractor is a $2$-cycle for any value of alpha, excluding very large alpha.  Indeed, at large $\alpha$, the dominant attractor is actually an independent set fixed point (each occupied site is isolated). 
\subsection{Additional plots for $K=2$ and $d\geq 3$}
\begin{figure}[p]
    \centering

    \begin{minipage}{\columnwidth}
        \centering
        \includegraphics[width=0.7\columnwidth]{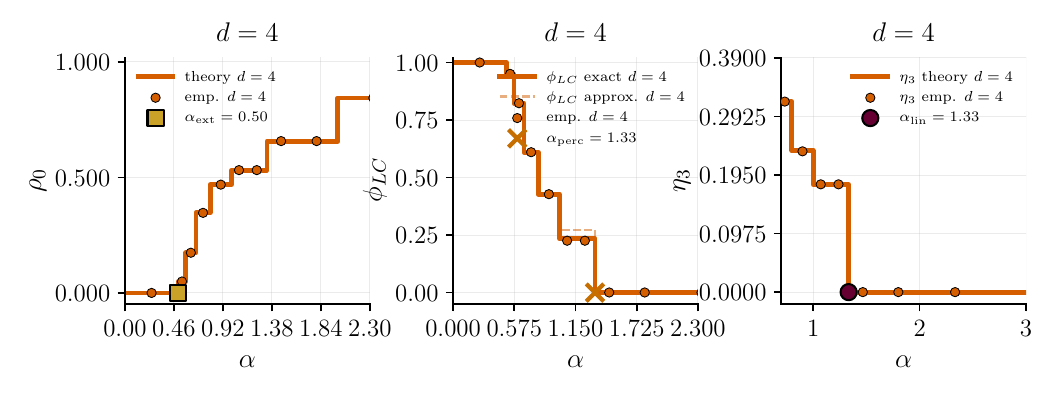}
    \end{minipage}

    \begin{minipage}{\columnwidth}
        \centering
        \includegraphics[width=0.7\columnwidth]{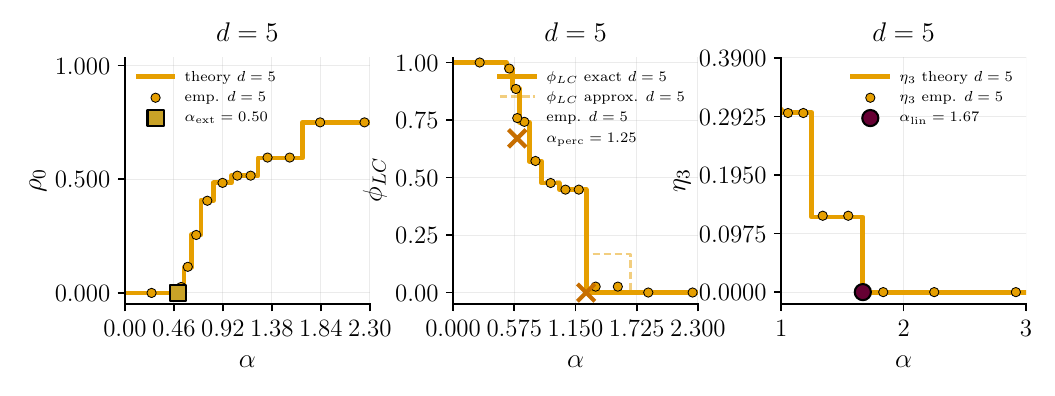}
    \end{minipage}

    \begin{minipage}{\columnwidth}
        \centering
        \includegraphics[width=0.7\columnwidth]{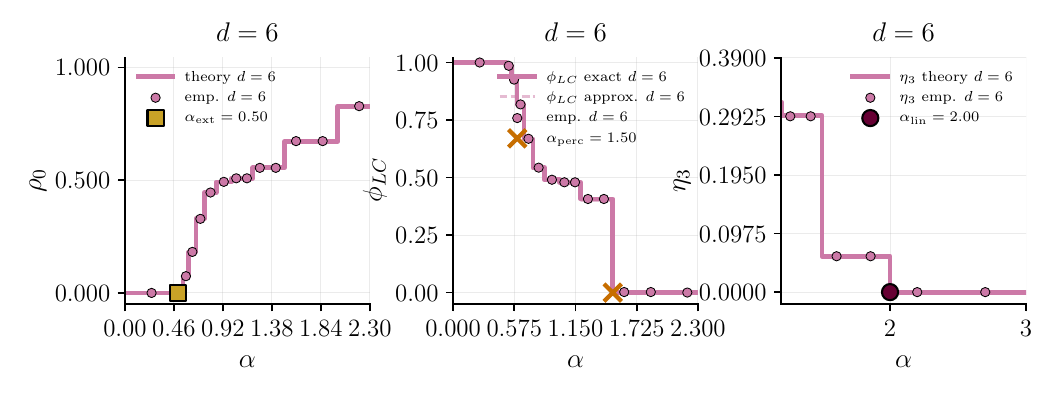}
    \end{minipage}

    \begin{minipage}{\columnwidth}
        \centering
        \includegraphics[width=0.7\columnwidth]{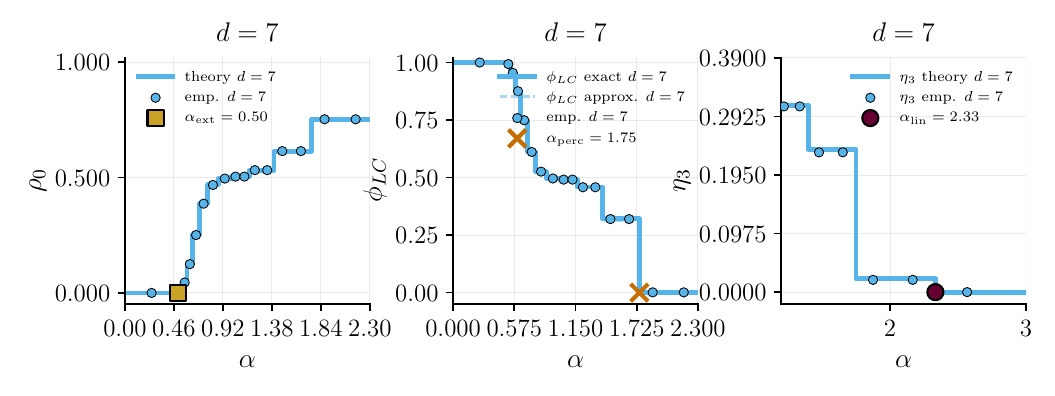}
    \end{minipage}

    \begin{minipage}{\columnwidth}
        \centering
        \includegraphics[width=0.7\columnwidth]{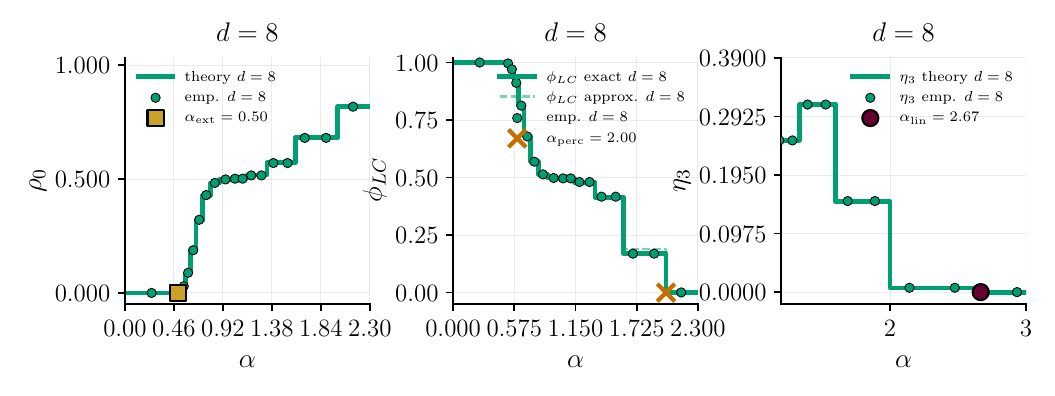}
    \end{minipage}

    \caption{
    Comparison of the numerical results with theory for $K=2$ at increasing graph degree $d$. We report $\phi_{LC}$, $\rho_0$, $\eta_{3}$. $\phi_{LC}$ is computed both with the exact formula as in Eq.\eqref{E24}, and with the approximation. 
    }
    \label{all_d_K_2}
\end{figure}

We show in Figure~\ref{all_d_K_2} the comparison between numerical simulations and theory for higher $d$. We show also the value of $\phi_{LC}$ obtained with the approximation discussed in Appendix~\ref{Percolation_Exact}, and compare it with the exact result. This supports the claim that the approximation works very well for almost all $d$, as the differences between approximation and exact theory are roughly for most alpha of order $10^{-2}$. Thus, there is a very good match between the approximated formula and the exact one for $\phi_{LC}$. 
\end{document}